\documentclass[aip,reprint]{revtex4-1}
\usepackage{graphicx}
\usepackage{epsfig}
\usepackage{subfigure} 
\usepackage{amssymb,amsmath}
\usepackage{overpic}
\usepackage{epstopdf}
\usepackage[usenames,dvipsnames,svgnames,table]{xcolor}

\usepackage[pdftex,bookmarks,colorlinks]{hyperref}

\newcommand{\be}{\begin{equation}}
\newcommand{\ee}{\end{equation}}
\newcommand{\bea}{\begin{eqnarray}}
\newcommand{\eea}{\end{eqnarray}}

\draft 

\begin{document}


\title{Geometry of flexible filament cohesion:  Better contact through twist?} 



\author{Luis Cajamarca}
\email[]{lcajamar@physics.umass.edu}
\affiliation{Department of Physics, University of Massachusetts, Amherst, Massachusetts 01003, USA}

\author{Gregory M. Grason}
\email[]{grason@mail.pse.umass.edu}
\affiliation{Department of Polymer Science, University of Massachusetts, Amherst, Massachusetts 01003, USA}



\begin{abstract}
Cohesive interactions between filamentous molecules have broad implications for a range of biological and synthetic materials.  While long-standing theoretical approaches have addressed the problem of inter-filament forces from the limit of infinitely rigid rods, the ability of flexible filaments to deform intra-filament shape in response to changes in inter-filament geometry has a profound affect on the nature of cohesive interactions. In this paper, we study two theoretical models of inter-filament cohesion in the opposite limit, in which filaments are sufficiently flexible to maintain cohesive contact along their contours, and address, in particular, the role played by helical-interfilament geometry in defining interactions.  Specifically, we study models of featureless, tubular filaments interacting via 1) pair-wise Lennard-Jones (LJ) interactions between surface elements and 2) depletion-induced filament binding stabilized by electrostatic surface repulsion.  Analysis of these models reveals a universal preference for cohesive filament interactions for non-zero helical skew, and further, that in the asymptotic limit of vanishing interaction range relative to filament diameter, the skew-dependence of cohesion approaches a geometrically defined limit described purely by the close-packing geometry of twisted tubular filaments.  We further analyze non-universal features of the skew-dependence of cohesion at small-twist for both potentials, and argue that in the LJ model the pair-wise surface attraction generically destabilizes parallel filaments, while in the second model, pair-wise electrostatic repulsion in combination with non-pairwise additivity of depletion leads to a meta-stable parallel state. 
\end{abstract}

\pacs{}

\maketitle 



%
%

%


\section{Introduction}
\label{sec:intro}
Assemblies of filamentous molecules driven by cohesive interactions continue to draw intense interest not only for the broad relevance to biology and nanostructured materials, but also due to the rich interplay between interactions and mechanics they exhibit.  Commonly studied biological examples include condensation of DNA\cite{adrian1992,podgornik1998,rau08,rau10}, filamentous actin\cite{ikawa2007,lau2009,lieleg2007,hosek04,needleman04,tang1996}, filamentous viruses\cite{tang2002} and protein fibrils\cite{serpell2000,weisel1987}, while synthetic examples include gelators\cite{douglas09}, supramolecular fibers\cite{meijer01} as well as carbon nanotube ropes \cite{liang2005,terrones1997,sola2010}.  As the emergent properties of assembly motifs like protein filament bundles and nanotube ropes depend not only on intrinsic properties of single filaments, but also, on the properties of inter-filament forces, developing a theoretical understanding of cohesion between filamentary objects has been a long standing goal.  

\begin{figure}[htbp]
\begin{center}
\includegraphics[scale=0.35]{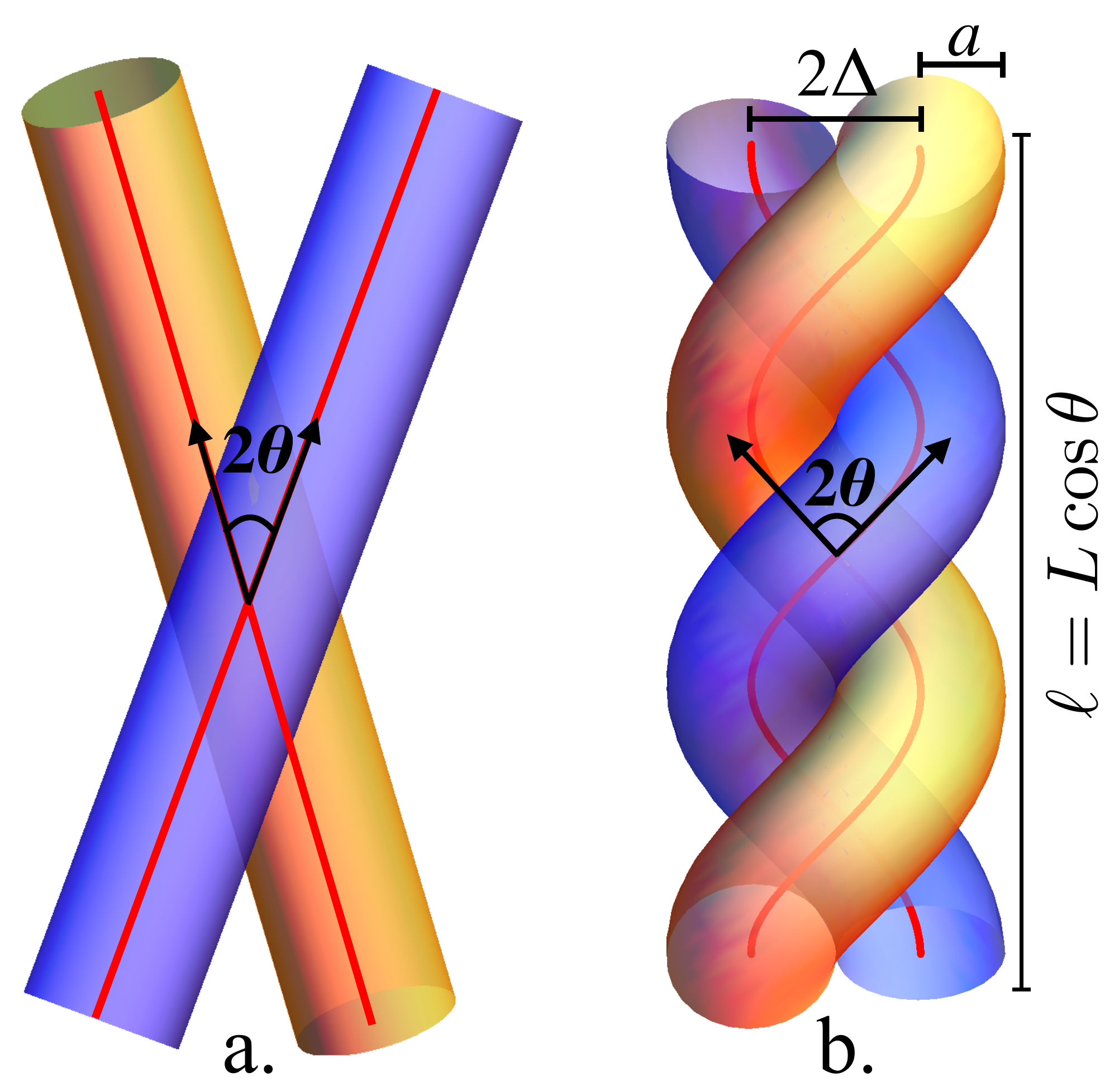}
\caption{(a) Rigid, skewed filaments, with non-optimal length of contact. (b) Model of twisted flexible tubes.  The radius of the tubes is $a$ and the center-to-center separation between centerlines is $2\Delta$. The vertical height decreases with non-zero skew (twist) via $\ell= L \cos\theta$.  The local skew angle between the filaments is $2\theta$.}
\label{fig:double_helix}
\end{center}
\end{figure}

Even in the simplest models of interacting filaments, the extreme length to radius aspect ratio, $L/a \gg 1$, implies a strong dependence of inter-filament forces on orientation, in particular, on the inter-filament skew angle, $\theta$ (see Fig.~\ref{fig:double_helix}).  For example, van der Waals attractions between perfectly rigid and skewed cylinders has been studied theoretically~\cite{adrian1972, mitchell1973,ohshima1996,harries1998}, showing that for a given separation, cohesive energy {\it decreases} with skew angle as $\sim 1/\sin \theta$.  This intuitive result simply derives from the fact that parallel configurations maximize the effective length of ``contact" between straight filaments, and even small rotations of rigid filaments move distant regions out of cohesive contact Fig.~\ref{fig:double_helix}(a).  

While the skew-dependence of pair-wise filament interactions has been well-studied theoretically in the limit of perfect rigidity, the purpose of the present study is to revisit the angle-dependence of cohesion for the case of {\it flexible} filaments
, which we show exhibits a critically different behavior.  The extreme aspect ratio of filaments, generically implies that in their optimal geometry, skewed pairs of cohesively interacting filaments are unlikely to remain straight  and instead are typically deformed by cohesive interactions~\cite{vanGestel2007}. Specifically, we consider a class of structures where filaments maintain cohesive contact along their contour length:  the {\it double helical} configuration (Fig.~\ref{fig:double_helix}(b)).   Note that for a fixed non-zero skew,  long filament pairs gain cohesive energy in proportion to $L$, while the mechanical cost for filament bending generically vanishes rapidly with radius (e.g. as $\sim a^4$ for the isotropic elastic beams).  These dimensional arguments alone imply that the double-helical pair configuration is a more relevant geometry than ``crossed cylinders" for describing minimal-energy states of cohesive pairs of sufficiently long and slender filaments.  

Motivated by these simple observations, we study two theoretical models of cohesive interaction between featureless, tubular filaments in order to address three basic and unanswered questions.  First, what is the angle dependence of cohesive interactions between double-helically wound filament pairs?  What determines the optimal skew geometry of cohesive contact?  Finally, how do cohesive interactions between bound pairs of flexible filaments affect the stability of the parallel configuration?  In the first model, we study the cohesive interactions mediated by a pair-wise attraction between surface elements of opposing filaments, whose distance-dependence is modeled by a Lennard-Jones (LJ) potential, characterized by an attractive well at separation $\sigma$.   This potential provides a simple, albeit approximate, means of modeling long-range van der Waals forces between tubular filaments, and more importantly for present purposes, its analytical simplicity allows us to characterize the full skew-dependence of inter-filament cohesion and classify its behavior purely in terms of the ratio of range of interactions to the filament radius, $\sigma/a$.  The second model considers the cohesion between osmotically-condensed filaments (depletion driven) stabilized by screened electrostatic repulsion between the charged filament surfaces, a potential relevant to numerous experimental studies of biofilaments condensed in crowded suspensions of macromolecular depletants\cite{hosek04,lau2009,needleman04}.  Previous approaches to the skew-dependent cohesion of flexible filaments have relied on heuristic and purely geometric descriptions of interactions, such as the assumption that the strength of inter-filament cohesion can be traced directly to the length of certain lines of ``contact" between twisted filaments~\cite{vanGestel2007}.  In contrast, the present study relies only on direct evaluation of microscopic models of interaction between twisted filament surfaces, allowing us to directly assess the range of validity of any such heuristic model of interfilament cohesion.

Based on our analysis, we find that despite the clear distinctions in microscopic mechanisms of surface cohesion and the pair-wise additivity of surface interactions considered, both models exhibit a generic preference of interactions for twist:  cohesive energy in the filament pair is maximized at non-zero skew angle.  It is important to distinguish this result for preferred spontaneous twist of achiral filament pairs, from the well-studied intrinsic preference for twist in models of helically-patterned, or chiral, filaments \cite{samulski1977,kornyshev2000,kornyshev2007,grason2007,cortini2011}.  For the case of purely pair-wise and attractive interactions among surface elements of featureless filaments (such as the LJ model), we show further that cohesive interactions generically destabilize parallel filaments to at least a small degree of inter-filament skew, whose value derives from a balance between the cohesive inter-filament torque and the mechanical forces of filament bending.  We argue that certain features of the skew-dependence of interactions derive from universal features of the contact geometry of twisted tubular filaments.  Specifically, independent of the mechanism of the interaction, we show that in the limit of very short-range surface interactions, the optimal cohesive geometry is the one that maximizes the length of {\it lines of contact} between the surfaces of close-packing helical tubes, a well-defined and purely geometric state.

The remainder of this paper is organized as follows:  an introduction to the geometry of twisted filament pairs and an overview of the contact geometry of twisted tubes is given in Sec.\ref{sec:model}. In Secs.\ref{sec:interactions_LJ} and \ref{sec:interactions_DHD}, we introduce and analyze the two models for the molecular interactions between filaments.  In Sec.\ref{sec:interactions_LJ} we study the LJ cohesive potential first focusing on the dependence of optimal skew angle on the ratio of range of the interaction to filament radius.  We then focus on the stability of parallel filaments to non-zero skew, based on the LJ model of filament attraction as well as the mechanical cost for filament deformation. A parallel analysis is carried out in Sec.\ref{sec:interactions_DHD} for the model of osmotically-condensed/electrostatically-stabilized (OCES) filament pairs, where we investigate how skew-dependent interactions depend on the ranges and relative strengths of the respective electrostatic and depletion forces between filaments. In Sec.\ref{sec:disc} we discuss both the universal and interaction-specific features of the skew-dependence of filament interactions revealed by our study, as well as the implications for specific filamentous systems.  Finally, we provide details on the calculation of depletion-attraction between twisted tubular filaments in Appendix \ref{app:depletion}.

\section{Geometry and contact of helical filament pairs}
\label{sec:model}

\begin{figure}[htbp]
\begin{center}
\includegraphics[scale=0.28]{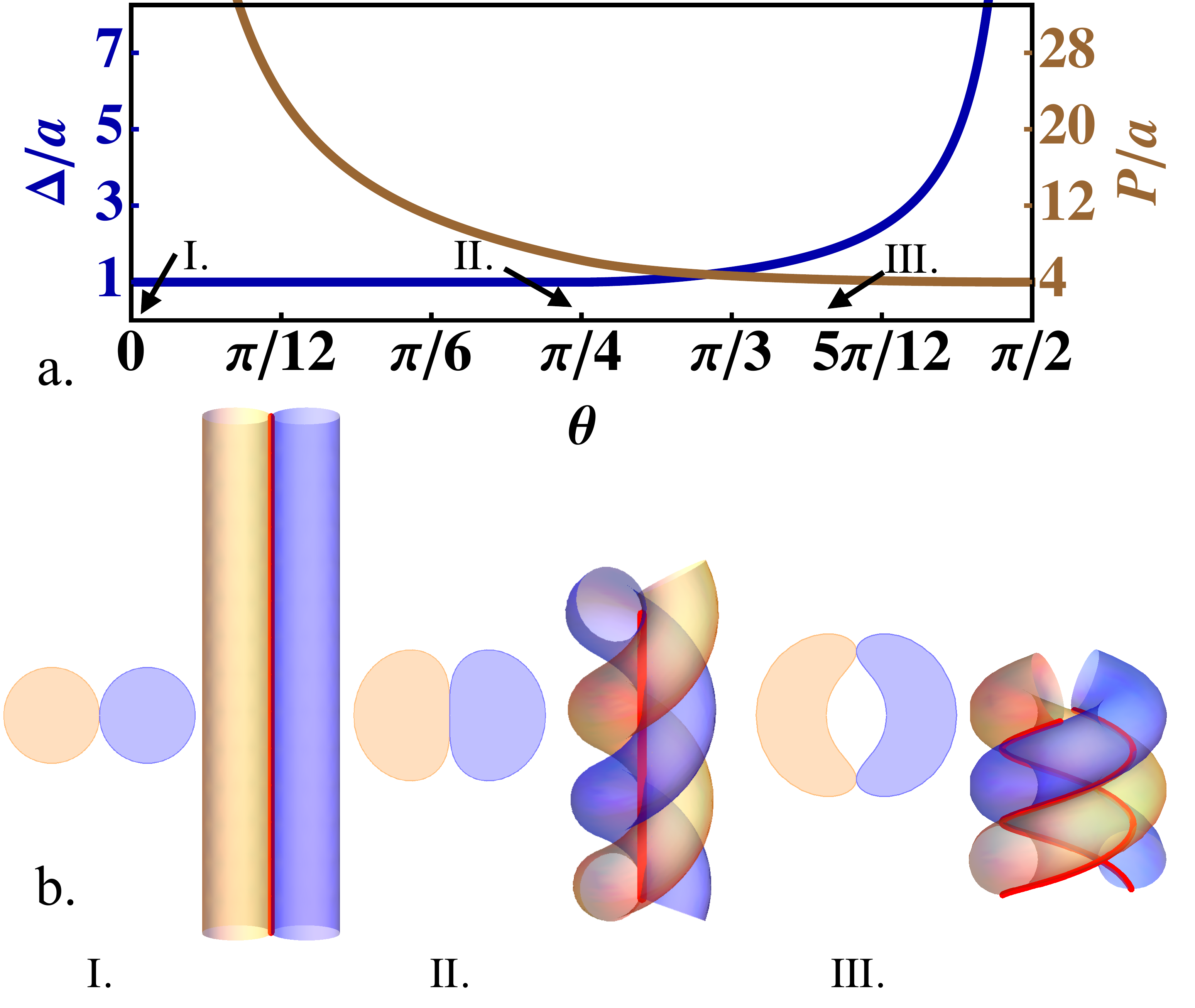}
\caption{(a). Non-linear behavior of pitch, $P/a$, and spacing, $\Delta/a$, for close-packed double helices. (b). Top (cross-sectional) and front view of double-helix packing for three different helical angles: I. $\theta=0$, II. $\theta=\pi/4$ and III. $\theta=7\pi/18$. The red lines correspond to the lines of contact. Notice the opening in between the filaments for  the cross section in III, where $\theta> \theta_c$. Both the height and the cross sections are drawn to scale.}
\label{fig:delta_pitch}
\end{center}
\end{figure}

In this section we introduce the geometry of twisted tubes used to describe interactions between featureless filaments, the surfaces of which are both homogeneous and isotropic in cross section.  Specifically, we consider filaments of radius $a$ and length $L \gg a$, and review geometric properties of interfilament contact in ``close packed" helical tubes.  

We treat filaments as ``flexible tubes" possessing a circular cross section perpendicular to the central curve, or tube axis, at each point along its length, which is sometimes referred as the ``tube" picture of molecular filaments~\cite{banavar2003}.  The curves describing a twisted pair of filaments are helices, which we denote as ${\bf r}_+(z)$ and ${\bf r}_-(z)$ where $\pm$ simply labels each filament and $z$ is the vertical height along the pitch axis.  Denoting the center-to-center spacing between the axes as $ 2\Delta$ and the pitch as $P= 2 \pi/ \Omega$\footnote{$\Omega$ is usually referred to as the \emph{rate of twist}.} these curves are simply
\begin{equation}
\label{eq: rpm}
{\bf r}_\pm (z) =\pm \Delta \big[ \cos( \Omega z) \hat{x} + \sin( \Omega z) \hat{y} \big] +z ~\hat{z} .
\end{equation}
The pitch and radii of the tube axes define the {\it helical angle}, $\theta$, through
\begin{equation}
\tan \theta = \Omega \Delta ,
\label{eq:tantheta}
\end{equation}
where according to this definition, the local skew angle between the filament pair is $2 \theta$.  Finally, we describe the points on the surface of the tubular filaments in terms of the Frenet frame $\{ {\bf T}_\pm, {\bf N}_\pm, {\bf B}_\pm \}$, the tangent, normal and binormal to each curve \cite{kamien2002},
\begin{equation}
{\bf T}_\pm = \frac{ \Omega ~ \hat{z} \times {\bf r}_{\pm} (z) + \hat{z} } {\sqrt{1+ (\Omega \Delta)^2} }; {\bf N}_\pm = \mp  \Big[ \cos( \Omega z) \hat{x} + \sin( \Omega z) \hat{y} \Big] ,
\label{eq:T_and_N}
\end{equation}
and ${\bf B}_\pm  = {\bf T}_{\pm} \times  {\bf N}_{\pm}$.  Surface elements of each filament are labeled by the coordinate pair $(z_\pm, \phi_\pm)$, where the latter coordinate labels the angular location with respect to the tube center, whose position is described by the function ${\bf R}_\pm (z_\pm, \phi_\pm)$,
\begin{equation}
\label{eq: Rpm}
{\bf R}_\pm (z_\pm, \phi_\pm) = {\bf r}_\pm (z_\pm) + a \big( \cos \phi_\pm ~ {\bf N}_\pm + \sin\phi_\pm ~ {\bf B}_\pm \big) .
\end{equation}
Below, we consider models of cohesion deriving, in part, from the summation of interactions among surface elements on opposing filaments.  Given the parameterization of eq.~\eqref{eq: Rpm}, area elements are related to surface coordinates via a metric $d A_{\pm} = \sqrt{g_{\pm} } d z_\pm d \phi_\pm$ where
\begin{equation}
\sqrt{g_\pm} = 
|\partial_{z_{\pm}} {\bf R}_{\pm} \times   \partial_{\phi_{\pm}} {\bf R}_{\pm}|= a \sec \theta \big[ 1 - (a/\Delta) \sin^2 \theta \cos \phi_{\pm} \big]  .
\end{equation}
Notably, the factor of $\sec \theta$ relates a length element along the pitch axis $dz$ to the arc length of the tube axis.  The fixed contour length of filaments implies that the height of the pair along the pitch axis, $\int dz_{\pm} =\ell$, must contract for non-zero twist according to $\ell = L \cos \theta$ (see Fig.~\ref{fig:double_helix}(b)).  

Before proceeding to analyze cohesive interactions for helical filament pairs, we first review some geometric aspects of inter-filament contact for close-packed helical tubes.  The geometric constraints of tube packing provide a clear illustration of the non-local nature of contact in multi-filament structures.  Moreover, we show below that the close-packing geometry of twisted tube pairs encodes certain generic features of the cohesive energy landscape for attractive filaments.   

Considerations of non-overlap between filaments in ``n-ply" geometries have been studied in detail, first by Neukirch and van der Heijden\cite{heijden2002} and more recently by Bohr and Olsen\cite{olsen2010}.  Close-packed configurations of double-helical filaments (or ``2-plies"), refer to configurations where the distance of closest approach between the axes of opposing tubes is identically equal to the diameter $2 a$, a condition which constrains the relationship between $\Delta$, $P$ and $a$.  Defining the distance between the $+$ and $-$ tubes offset by a vertical distance $z$ as $\delta(z)\equiv|{\bf r}_+(z_0+z) - {\bf r}_-(z_0)|$ from eq.~\eqref{eq: rpm} we have
\begin{equation}
\delta^2(z) = 2 \Delta^2 \big[1+ \cos(\Omega z) \big] + z^2 .
\end{equation}
Lines of contact between the tubes are defined as solutions to
\begin{equation}
\label{eq: doca}
\partial_z \delta^2(z) |_{z=z_*} =2 z_* - 2 \Omega \Delta^2 \sin (\Omega z_*) =0 ,
\end{equation}
which requires that points separated by $z_*$ are locally the distance of closest approach between the tube axes.  For a given $\Omega$ and $\Delta$, there are, in general, multiple roots to the transcendental equation of eq.~\eqref{eq: doca}, and non-overlap between the filament pair requires that $\delta(z_*) \geq 2 a$ for all solutions, while {\it close-packing} requires that at least one solution saturates the inequality (i.e. the pair can be brought no closer without overlap).  

Straightforward analysis of eq.~\eqref{eq: doca} shows that for small twist $|\Omega \Delta| \leq 1$, or $\theta \leq \pi/4$, there is only a solution for $z_*=0$ (i.e. contact occurs at the same vertical position).  Hence, for $\theta \leq \theta_c = \pi/4$ close-packing is described by $\Delta = a$ and $\Omega = a^{-1} \tan \theta$.  Above a critical helical angle, $\theta_c = \pi/4$, a bifurcation occurs and the solution to eq.~\eqref{eq: doca} splits into two lines of contact, occurring out of the plane (i.e. $z_* \neq 0$).  For $\tan \theta \gtrsim 1$, it is straightforward to show $\Omega z_* \simeq \pm2 \sqrt{2} \sqrt{ \theta - \theta_c}$.  Further, above the critical angle $z=0$ becomes a local {\it maximum}, implying that $\Delta > a$.  For the asymptotic case at large helical angles $|\Omega \Delta| \gg 1$, it is straightforward to show that contact is described by the conditions $ \Omega z_* \simeq \pm \pi$, $P \simeq 4 a$ and $\Delta = 4 a \tan \theta$.  

The evolution of close-packed geometry of double helical tubes with increased twist is shown in Fig.~\ref{fig:delta_pitch}. For small twist ($\theta \leq \pi/4$) there is a single line of contact between tubes which threads along the helical axis ($x=y=0$)~\footnote{This definition of contact line, which measures the length of curves of inter-surface contact in the {\it deformed} and twisted geometries (the vertical line threaded through the center of the double helix) differs from the notion of contact in ref. \cite{vanGestel2007}, which considers the length of curve contacting points when the twisted filaments are unbent into straight configurations.  Notably, the line of contact in the twisted configuration {\it decreases} with twist, while ``unwrapped'' line of contact {\it increases} with twist.}.  For large twist  ($\theta \geq \pi/4$) a given tube makes contact with the opposing tube at {\it two points}, in the helical turns above and below a given point.  At the critical angle $\theta_c = \pi/4$, these two distinct contact geometries merge, indicating a unique and broadly distributed neighborhood of ``near-contact" between opposing tubes, $\delta^2(z) - (2 a)^2 \simeq z^4 a^{-2} /12$. Crudely, we may think that this critical geometry possesses simultaneously {\it three lines of contact}, both the central line of contact occurring for small twist as well as the two helical lines describing out-of-plane contact at large twist.  In the analysis of interaction energy between tubular filaments, we show that the existence of a critical point in the contact geometry of close-packed flexible tubes at $\theta_C = \pi/4$ has important and universal consequences for the twist-dependence of cohesive energy.

\section{Case I: van der Waals attraction}
\label{sec:interactions_LJ}

\begin{figure}
\begin{center}
\includegraphics[scale=0.35]{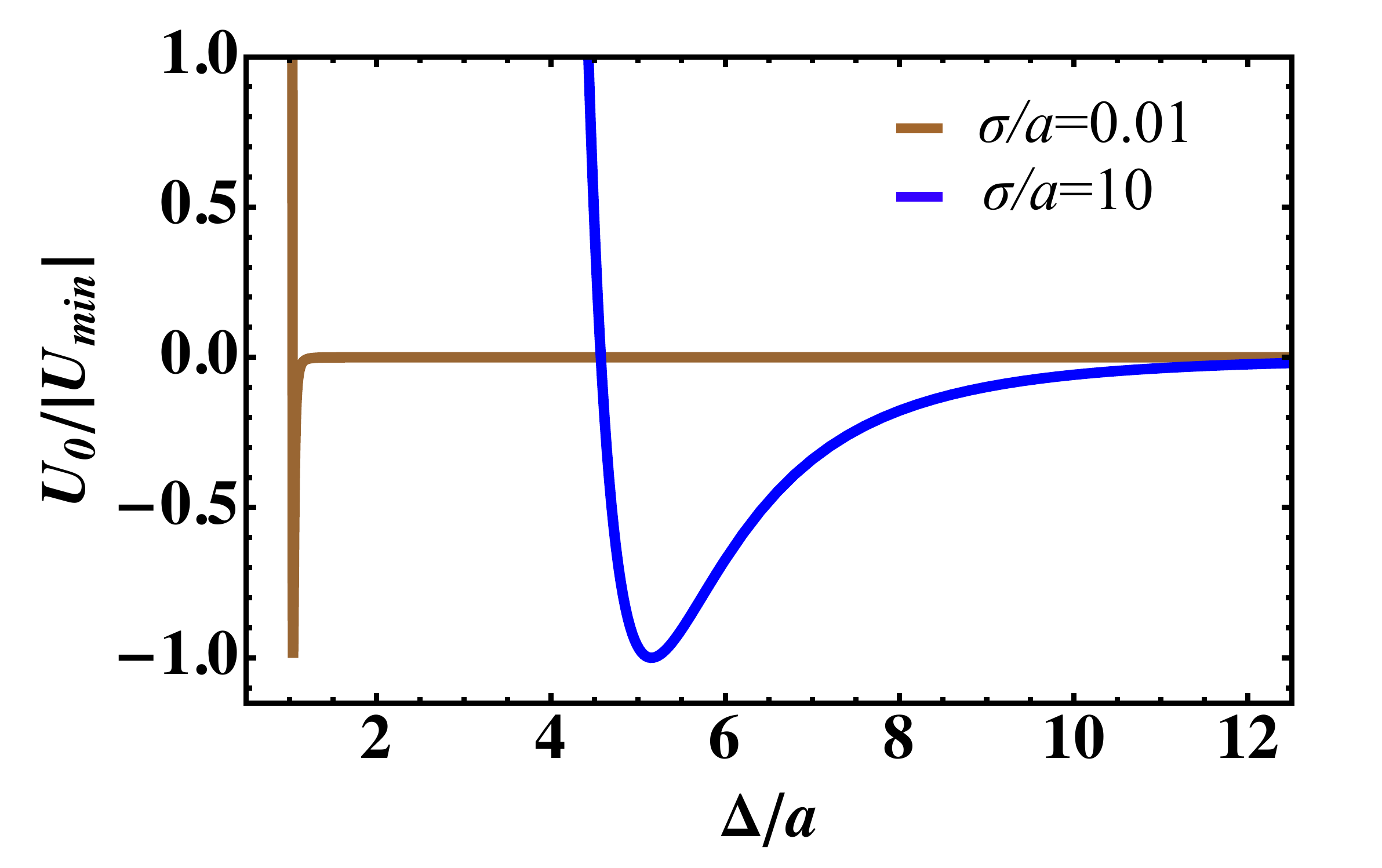}
\caption{Normalized LJ interaction energy as a function of dimensionless separation $\Delta/a$ for $\theta=0$. The {\it sticky tubes} limit corresponds to $\sigma/a=0.01$ whereas the {\it sticky threads}  to $\sigma/a=10$.}
\label{fig:LJ_both}
\end{center}
\end{figure} 

In this section we analyze the $\theta$-dependence of the double-helical filament pairs interacting via a pair-wise attraction between surface elements on opposing filaments.   Here, we consider pair-wise attractions between tubular surface elements modeled by LJ-type potential, whose long range $-r^{-6}$ term models non-retarded van der Waals forces,
\begin{equation}
u_{LJ}(r) = \epsilon\bigg[ \Big( \frac{ \sigma}{r} \Big)^{12} - 2 \Big( \frac{ \sigma}{r} \Big)^{6} \bigg] ,
\end{equation}
where $\sigma$ and $\epsilon$ denote the radial location and depth, respectively, of the attractive potential minimum.   For a given helical angle and filament spacing, $\Delta$, we compute the total interaction energy from the double surface integration of the pair-wise interactions,
\begin{equation}
U_{LJ} (\theta, \Delta) = \int dA_+ \int dA_- ~u_{LJ} \big(|{\bf R}_+- {\bf R}_- |\big) .
\label{eq:ulj_1}
\end{equation}
In this study we focus on interactions between filaments much longer than the interaction range ($L\gg \sigma$).  Hence, for a given fixed position, ${\bf R}_+$, we approximate the integral over the height coordinate of the $-$ filament by taking the upper and lower limits to $z_- \to \pm \infty$.  In this case, the interaction energy of elements on the $+$ filament with the surface of the $-$ are independent of height $z_+$.  Therefore, we arrive at the interaction energy per unit filament length,
\begin{multline} 
\label{eq: ULJ}
U_{LJ}(\theta, \Delta)/L = \cos \theta \int_0^{2 \pi} \sqrt{g_+} d \phi_+ \\
\times \int_0^{2 \pi}   \sqrt{g_-} d \phi_-  \int_{-\infty}^{\infty} dz_- ~u_{LJ}\big(|{\bf R}_+- {\bf R}_- |\big) .
\end{multline}
As our focus is on the skew-dependence of interactions, for a given angle $\theta$, we minimize interaction energy over spacing, defining $U_{LJ} (\theta) \equiv {\rm min}_{\Delta} [U_{LJ}(\theta, \Delta)]$.  Integrals and energy minimization are performed numerically.

The LJ potential is characterized by a single energy scale entering $\epsilon$, which has units of energy/(length)$^4$.  Therefore, up to rescaling by $\epsilon$, variations in inter-filament potentials are characterized a single parameter, $\sigma/a$, the ratio of surface interaction range to filament radius. Considering briefly the parallel filament ($\theta = 0 $) behavior, we highlight two limiting cases characterized by this ratio.  In the limit of $\sigma/a \to 0$ the inter-filament forces approach ``sticky tubes", whose interactions are negligible until surface elements of opposing filaments are in physical contact.  Hence, equilibrium spacing approaches $\Delta_0 \to a+0.4 \sigma$ for $\theta=0$, and expanding $|{\bf R}_+- {\bf R}_- |$ around $\phi_+=\phi_-=0$ in eq.~\eqref{eq: ULJ} (i.e. the Derjaguin approximation) we can show for cohesive tubes
\begin{equation}
\label{eq: U0tube}
\lim_{\sigma/a \to 0 } U_0/L \simeq - 4.48 a^{1/2} \sigma^{5/2} \epsilon
\end{equation}
where $U_0 \equiv U_{LJ} (\theta = 0)$ is the parallel tube interaction.  In the opposite limit, $\sigma/a \to \infty$, the range of filament interactions is characterized by $\sigma$ only, and we denote this as the ``sticky threads" limit.  For sticky threads, the distance dependence of the parallel configuration follows a ``5-11" potential\cite{grason2012,bruss2012,bruss2013}
\begin{equation}
\lim_{a/\sigma \to 0} U_{LJ} ( \theta =0 , \Delta)/L = 4\pi^2 a^2 \epsilon \bigg[\gamma_{12} \frac{\sigma^{12}}{\Delta^{11}} - 2 \gamma_{6} \frac{\sigma^6}{\Delta^5} \bigg] ,
\end{equation}
where $\gamma_n\equiv2^{-(n-1)} \int_{-\infty}^{\infty} du /(1+u^2)^{n/2}$.   In the thread limit, equilibrium spacing occurs at $\Delta_0 \simeq 0.474 \sigma$ and with a binding energy
\begin{equation}
\label{eq: U0thread}
\lim_{a/\sigma \to 0 } U_0/L \simeq - 66.57 a^2 \sigma \epsilon .
\end{equation}
The schematic behavior of $U_0$ for both limiting cases is shown in Fig.~\ref{fig:LJ_both}.

\subsection{Skew dependence of interactions and optimal angle}
\label{subsec:interactions_LJ} 
\begin{figure}
\begin{center}
\includegraphics[scale=0.29]{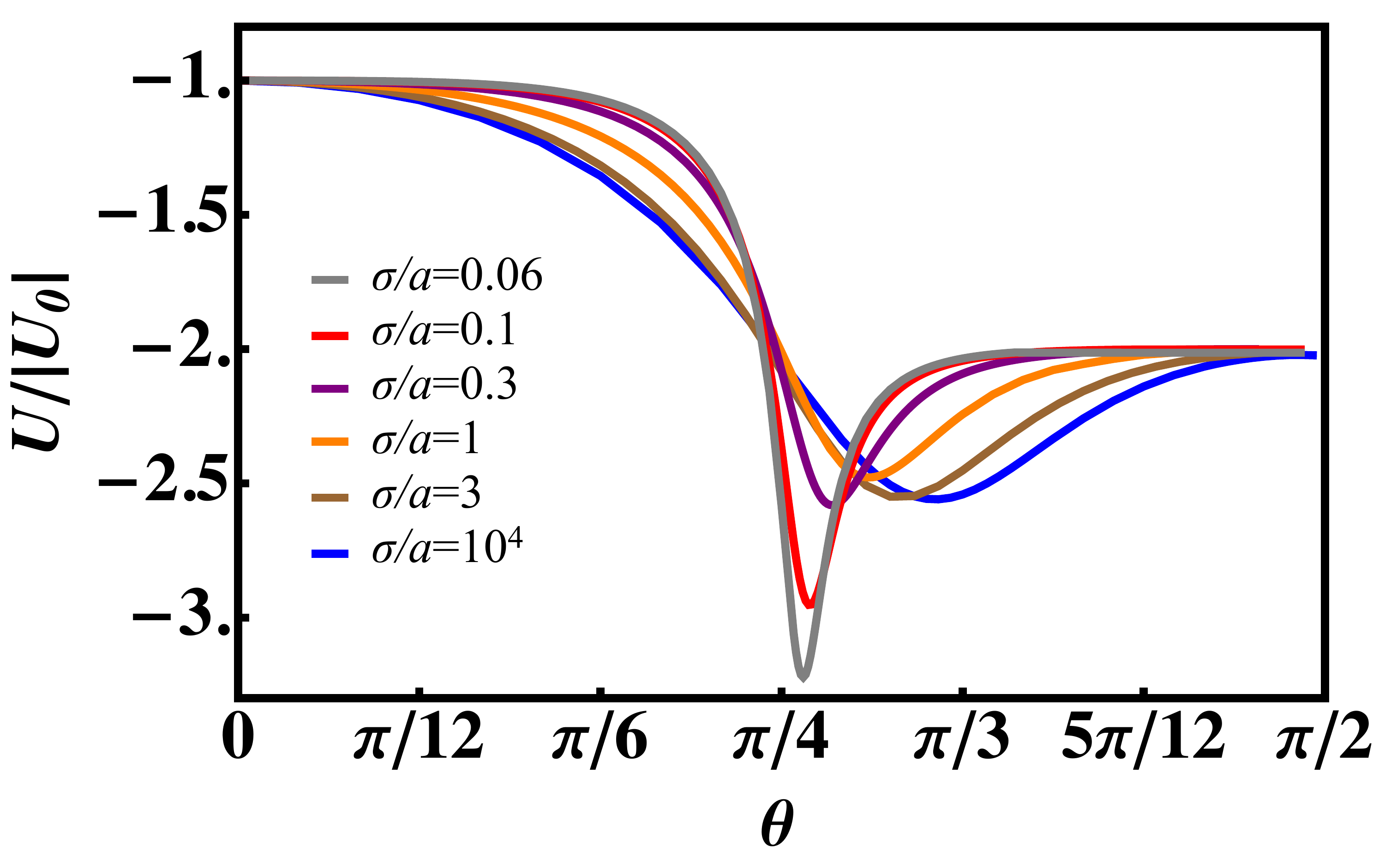}
\caption{Dimensionless energy $U/|U_0|$ as a function of helical angle $\theta$, optimized with respect to $\Delta$, for a  range of values of $\sigma/a.$ 
}
\label{fig:uvstheta}
\end{center}
\end{figure}

A plot of the normalized energy per unit length, \mbox{$U(\theta)/U_0$}, is shown in Fig.~\ref{fig:uvstheta} as a function of the helical angle $\theta$  for interaction ranges spanning from ``sticky tube" ($\sigma/a = 0.06$) to ``sticky thread" ($\sigma/a = 10^4$) limits. 

The angle dependence of interactions exhibits three common features for all ranges of interaction. First, the parallel configuration $\theta=0$ is an unstable local maximum of the cohesion energy, demonstrating that even infinitesimal skewing of parallel filaments enhances cohesive contact for the LJ surface interactions.  Second, the cohesion energy is optimal (most attractive) for a non-zero angle $\theta_m$, in the range $\pi/4 \leq \theta_m \leq 1.01$ ($45^\circ \leq \theta_m \leq 58^\circ$).  Finally, all interaction energies approach a second local maximum of $U \to 2 U_0$ as $\theta \to \pi/2$.  

Underlying this common behavior, are common features of inter-filament contact in the double-helical geometry.  The local maximum at $\theta = 0$ can be attributed to pair-wise attraction at long range between points on opposing filaments.  For parallel filaments at equilibrium spacing, all pairs are more distant than the closest separation $\Delta - 2a$ and sit within an attractive (negative) range of the potential.  Relative to this parallel state, double helical twist reduces the distance between points at a given vertical separation by reducing their distance in the xy plane.  For sufficiently small twist, this generically implies an increased number of points brought into the strongly attractive range of the pair-potential and a {\it increased} cohesive interaction, growing in proportion to $\sim \theta^2$.

\begin{figure}
\begin{center}
\includegraphics[scale=0.32]{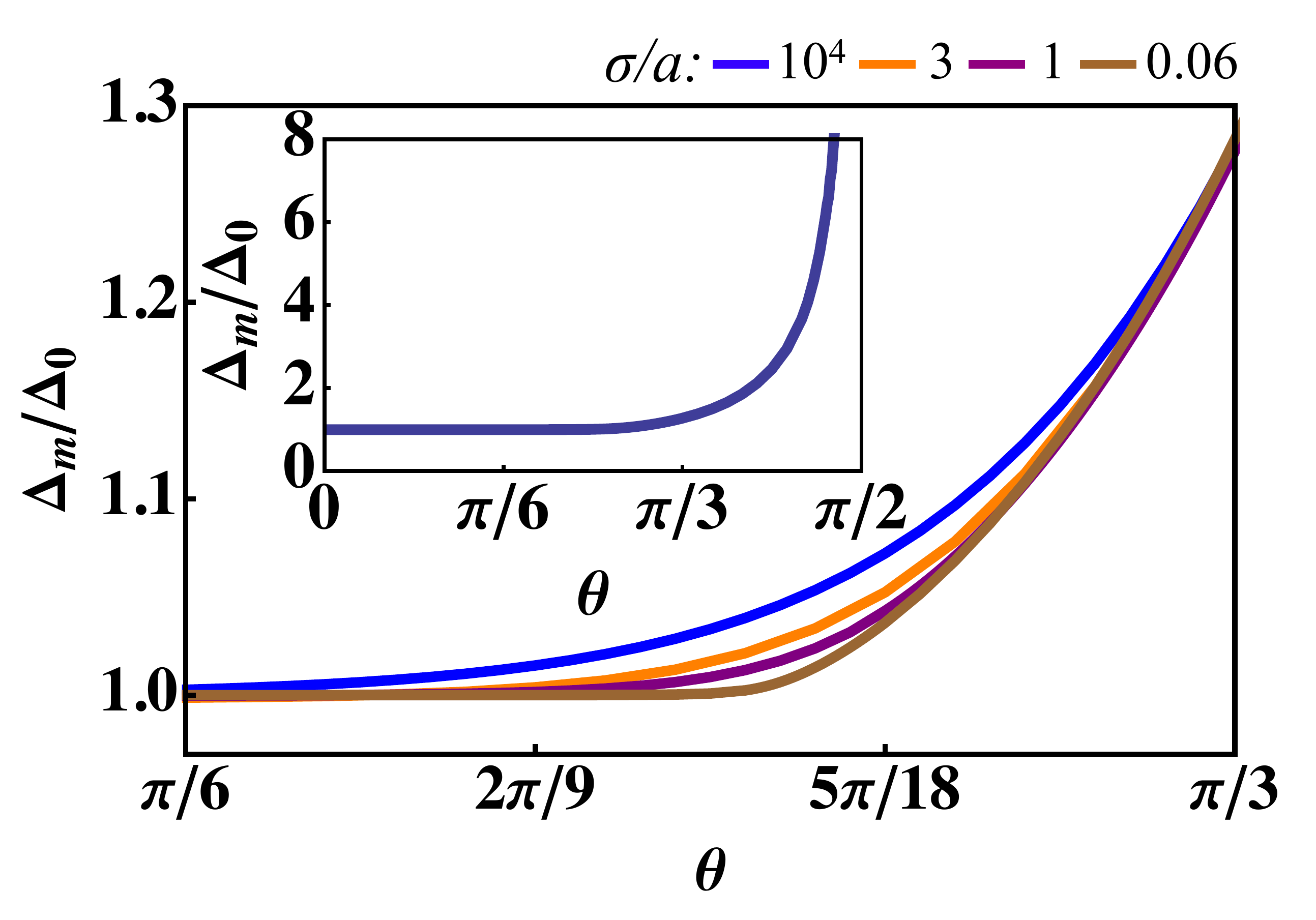}
\caption{Normalized equilibrium in-plane spacing $\Delta_m/\Delta_0$ as a function of $\theta$, for $\sigma/a=10^4,3,1$ and $0.06$. \emph{Inset:} The same plot for the full range of $\theta$. At this scale all values of $\sigma/a$ appear to be the same curve.}
\label{fig:delta_m}
\end{center}
\end{figure}

In Fig.~\ref{fig:delta_m} we plot the equilibrium in-plane spacing of filaments $\Delta_m$ as a function of $\theta$ for a range of $\sigma/a$.  For all interaction ranges, the optimal spacing is minimal for parallel filaments and ultimately diverges as $\theta \to \pi/2$, suggesting the pair opens a central void as the filament orientation tilts down into the xy plane at large twist angles.  The transition from $\Delta_m \approx \Delta_0$ at small angles $\theta \approx 0$ to the divergent in-plane spacing at large angles, $\Delta_m \sim 1/\tan \theta$, is a signature of the abrupt change of inter-filament contact discussed in Sec.~\ref{sec:model}. While contact between nearly parallel filaments occurs in the xy plane, for large angles the distance of closest approach occurs with two points on the neighbor filament {\it above} and {\it below} a point on the reference filament.  Hence, as the helical radius adjusts to $\Delta_m \sim 1/\tan \theta$, the interactions select the helical pitch so that as $\theta \to \pi/2$ the vertical spacing between ``stacked rings" approaches the optimal spacing of straight and parallel filaments, $\Delta_0$.  From these observations, it is straightforward to see that generic doubling of cohesion energy occurring in the large twist angle limit ($ \lim_{\theta \to \pi/2} U \to 2 U_0$) derives directly from doubling of the number contact lines between opposing filaments in this ``stacked rings" configuration as well as the filament straightening that occurs in asymptotic limit as $\Delta_m \to \infty$.

\begin{figure}
\begin{center}
\includegraphics[scale=0.3]{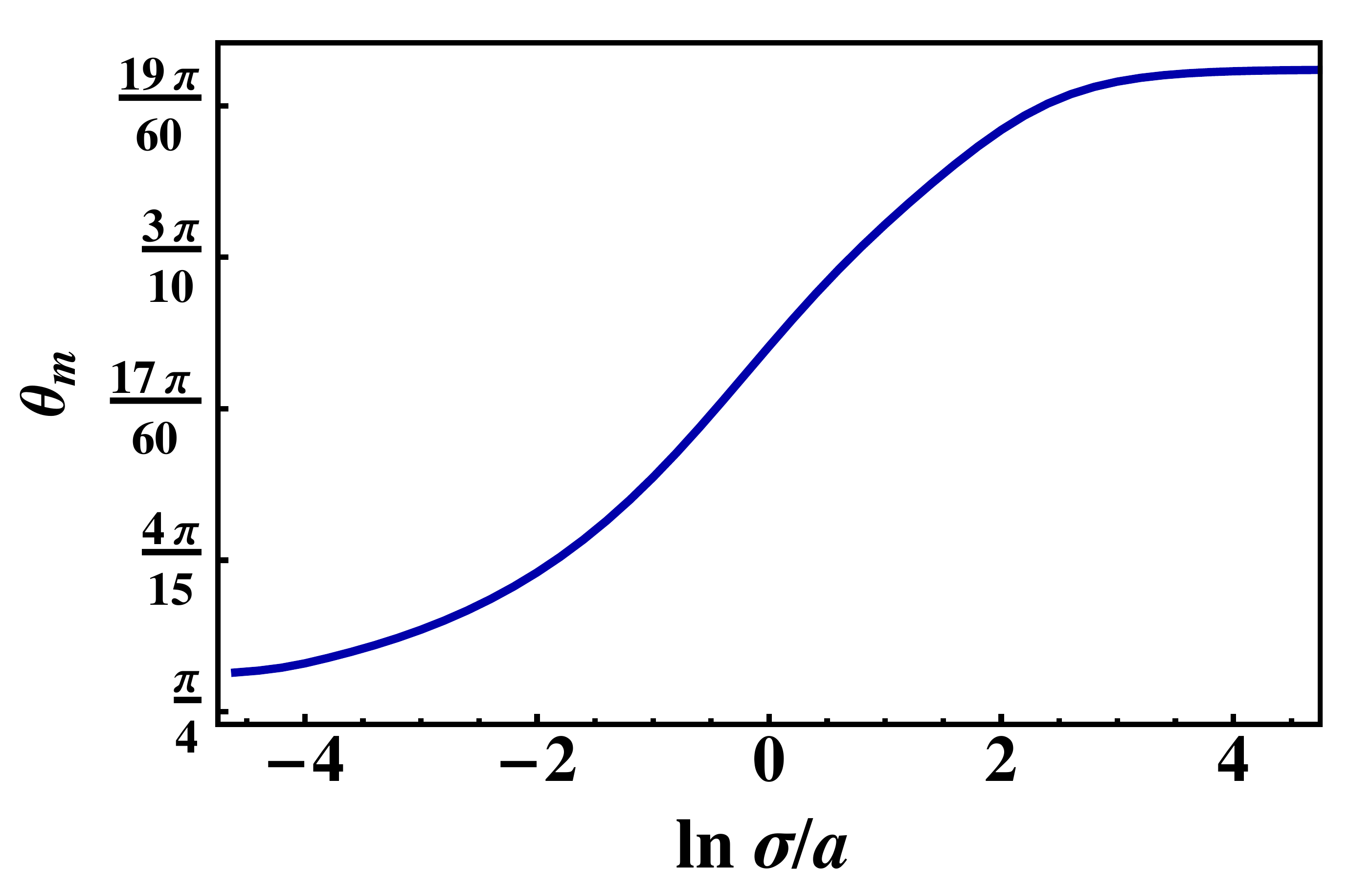}
\caption{Optimal cohesion angle plotted in terms of ratio of ranges of surface interaction to filament radius, showing asymptotic approaches to $\theta_m \simeq 1.01$ and $\theta_m = \pi/4$ in the respective ``sticky thread" and ``sticky tube" limits.}
\label{fig:thetamin}
\end{center}
\end{figure}

In Fig.~\ref{fig:thetamin} we plot the minimal-energy twist angle, $\theta_m$, as function of $\sigma/a$.  Notably, the optimal skew angle in the ``sticky tube" limit ($\sigma/a \ll 1$) approaches the critical angle of close-packed twisted tubes, i.e.  \mbox{$\lim_{\sigma/a \to 0} \theta_m \to \theta_c=\pi/2$} \footnote{Accuracy limitations of numerical integration of inter-filament potential limit the strict determination of the $\theta_m$ to $\ln (\sigma/a) \lesssim -4.5$}.  In the limit of vanishing interaction range, filaments only benefit from attractive interactions when opposing surfaces are in near-contact, and hence, we expect that the $\theta$-dependence of close-packed tubes also describes the optimal double-helical geometry of ``sticky tubes". Intuitively, we can understand the coincidence between the optimal cohesion geometry of sticky tubes and the critical-angle for close-packing in terms of the tendency to maximize the number of {\it lines of contact} between opposing filaments. At the critical angle, close-packed tubes transition from (at small angle) a single line of contact which threads along the pitch axis of the double helix to a configuration where a given filament is in contact with the opposing filament along two lines that connect consecutive and preceding helical turns (see Fig.~\ref{fig:delta_pitch}(b)).  Crudely, one may envision the transition of contact geometry at $\theta_c$ as the merger of {\it three lines of contact}, implying that the critical angle for close-packing is also the optimal angle for cohesion between sticky tubes.  Notably, this oversimplified view of the critical angle as a state of three contact lines is roughly consistent with observed depth of the minimal energy configuration relative to the parallel state for our LJ model, i.e. for the shortest interaction range reported $\sigma/a =0.06$ we find $U(\theta_m) \simeq 3.4 U_0$.  

In the limit $\sigma/a\gg1$,we find that the optimal angle approaches a value $\theta_m \simeq 1.01$ ($\theta_m \simeq58 ^\circ$). This angle characterizes the distinct geometry of optimally cohesive ``threads" whose much broader range of attraction between length elements allows attractive interactions to persist beyond the distance of closest approach between threads, and consequently, favors skew geometries which exceed the close-packing critical angle ($\pi/4$).  Hence, we can attribute the increase in $\theta_m$ with $\sigma/a$ to follow from the increase in the ratio of the distance of \emph{range} of pair-wise attractions between segments (proportional to $\sigma$) as compared to the equilibrium inter-filament spacing (roughly $\Delta_0 \approx a+ \sigma/2$).

\subsection{Bending and cohesive instability of parallel filaments}

The skew-dependence of cohesive interactions for LJ surface interactions shown in Fig.~\ref{fig:uvstheta} demonstrates that attractive interactions are locally maximal, and hence unstable, in the parallel configuration.  Focusing on the small-angle behavior, we find generically a decrease in energy, proportional to $\theta^2$.  We characterize the (in)stability of the parallel state in terms of the curvature, or second derivative, of the energy with respect to $\theta$,  
\begin{equation}
U(\theta) = U_0 + \frac{ U_2}{2} \theta^2 + O\big(\theta^4 \big) ,
\label{eq:theta_expand}
\end{equation}
where for the present case of LJ surface interactions, $d^2 U(\theta)/d \theta^2|_{\theta=0} = U_2<0 $.  The ratio $U_2/U_0$ characterizes the curvature of the interaction potential relative to the overall strength of attraction between filaments and we plot the dependence of $U_2/U_0$ on $\sigma/a$ in Fig.~\ref{fig:K2_plot}. Consistent with the ``flattening" of $U(\theta)$ in the ``sticky tube" limit ($\sigma/a \ll 1$) we find that the ratio $U_2/U_0$ decreases to zero as $\sigma/a \to 0$, exhibiting a linear power dependence on $\sigma/a$ in this limit.  We find the curvature grows in the ``thread" regime where $U_2/U_0$ becomes independent of $\sigma/a$, asymptotically approaching $U_2= 2U_0$ as $\sigma/a \to \infty$.  Combining this with the scaling-dependence of $U_0$ on $\sigma$ and $a$ [eqs.~\eqref{eq: U0tube} and \eqref{eq: U0thread}] in the asymptotic regimes of tube- and thread-like filament interactions we find
\begin{equation}
\label{eq:U_2_LJ}
|U_2|/L \sim \epsilon a^2 \sigma \times 
\left\{ \begin{array}{ll} 
(\sigma/a)^{5/2} , & \sigma/a \ll 1 \\  \\ 
1 , & \sigma/a \gg 1 \end{array} \right.
\end{equation}
highlighting the ratio of interaction range to radius as a key parameter dictating the cohesive drive for twist.  {We note in passing that the heuristic model of cohesion based on geometric contact between twisted tubes in ref. \cite{vanGestel2007} corresponds to the case $U_2 \propto - U_0$.  Based on the results of Fig.~\ref{fig:K2_plot}, we find, somewhat counterintuitively, that this heuristic assumption is consistent only with the thread limit ($\sigma \gg a$), while the tube limit ($\sigma \ll a$) shows a dramatic departure from this assumption, with the ratio of $U_2$ and $U_0$ strongly dependent on the ratio $\sigma/a$.

\begin{figure}
\begin{center}
\includegraphics[scale=0.3]{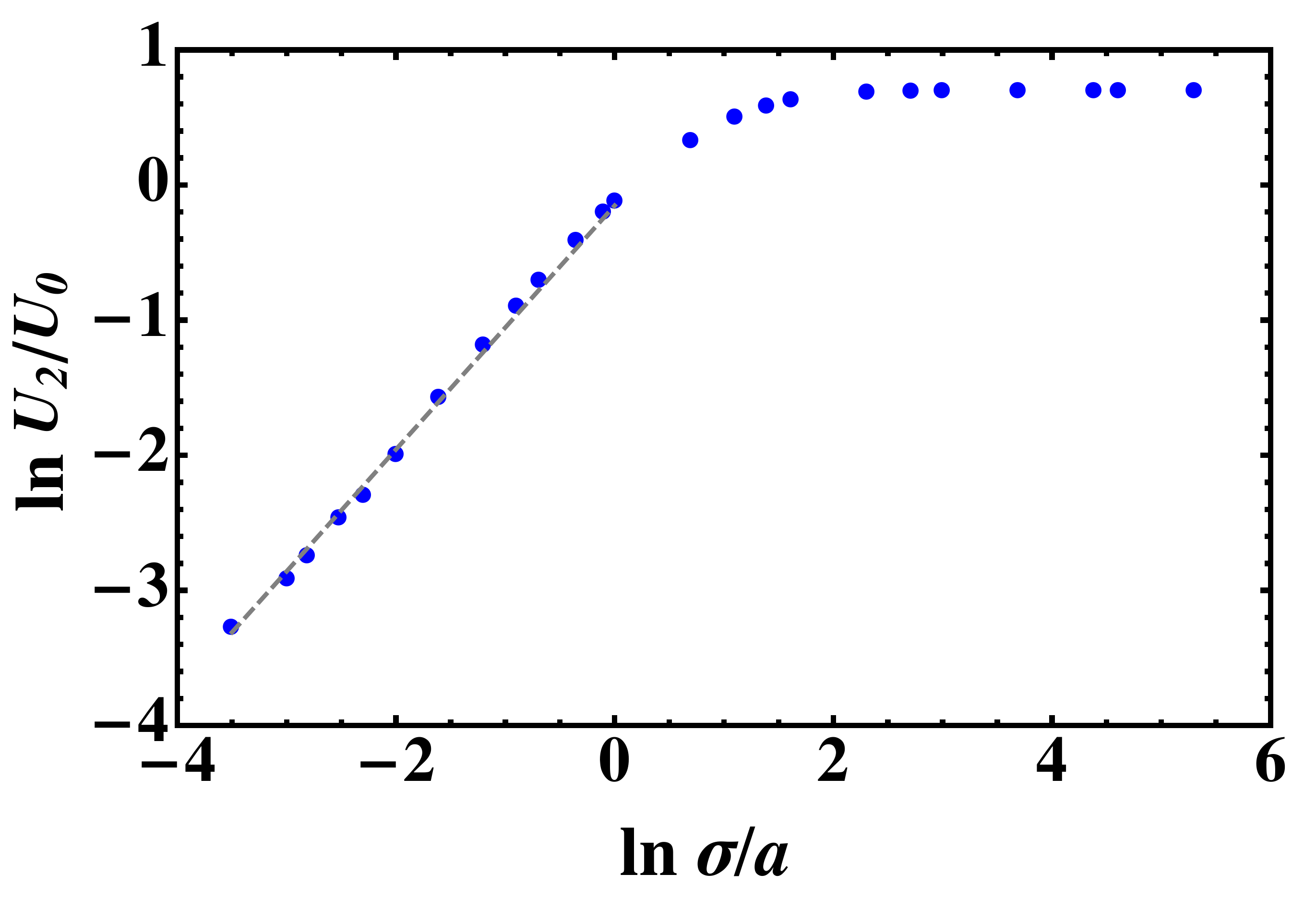}
\caption{Normalized curvature on angle-dependent potential $\ln (U_2/U_0)$ plotted in terms of ratio of interaction range to filament radius.  The slope of the dashed line implies $U_2/U_0 \sim \sigma/a$ in the limiting regime of ``sticky tube" interactions.}
\label{fig:K2_plot}
\end{center}
\end{figure}

The LJ model of surface-mediated pairwise cohesion of filaments implies a generic instability of parallel filaments to a double-helically twisted state.  We now consider the balance between cohesive LJ interactions and the mechanical costs of filament bending which determine the equilibrium skew angle of a filament pair.  The bending energy (per filament) is simply $B \kappa^2 L/2$, where $B$ is the bending stiffness of the filament and $\kappa = \Delta^{-1} \tan \theta \sin \theta$ is the curvature of the helical backbones.  Hence, the total mechanical cost of filament bending for small twist has the form,
\begin{equation}
E_{mech}/L = B \theta^4/\Delta^2 + O \big( \theta^6 \big) .
\end{equation}
Hence, the geometric dependence of curvature on twist, $\kappa \sim \theta^2$, implies that bending rigidity does not stabilize parallel filaments, and some cohesive interactions generically give rise to some degree of inter-filament skew.  Considering the limit of rigid filaments, where we expect the degree of inter-filament skew to be small, we find the equilibrium twist, $\theta_{eq}$ of filament pair,
\begin{equation}
\label{eq:theta_eq}
\theta_{eq} \simeq \frac{\Delta}{2} \sqrt{|U_2|/L B} .
\end{equation}
Because bending generates differential compression and extension throughout the cross section of the filament, bending stiffness is generically strongly dependent on filament radius. Here, we consider a simple model of an isotropic, elastic beam for which $B = \pi E a^4/4$, where $E$ is the material modulus of the filament \cite{landau1986}.  Combining the radius dependence of stiffness with the curvature of the cohesive potential, we find the asymptotic scaling of the equilibrium inter-filament skew on diameter and interaction range in the respective tube and thread regimes,
\begin{equation}
\theta_{eq} \sim \sqrt{ \epsilon \sigma / E } \times \left\{ 
\begin{array}{ll} 
(\sigma/a)^{5/4} , & \sigma/a \ll 1 \\ \\ 
\sigma/a, & \sigma/a \gg 1 \end{array} \right.
\label{eq:theta_cr}
\end{equation}
This analysis shows that although any filament pair well-modeled by van der Waals attraction between surfaces is unstable to some measure of interfilament twist, enhanced stiffness or small cohesive torques with large diameter filaments will lead to a marked reduction in $\theta_{eq}$ compared to smaller diameter filaments.  

\section{Case II: Osmotic attraction and electrostatic repulsion}
\label{sec:interactions_DHD} 
In this section, we consider the $\theta$-dependence of filament cohesion driven by two competing effects:  an \emph{osmotically-driven} attraction stabilized by an \emph{electrostatic} repulsion (OCES).  Our focus on this potential is motivated by the large body {\it in vitro} experimental study of biofilament (e.g. DNA, cytoskeletal filaments, filamentous viruses) condensation in solutions of inert polymers, which demonstrate that the respective repulsive and attractive interactions are independently controlled by ionic and osmotic solution properties.  In our model these competing effects are modeled within the context of the Debye-H\"uckel theory of screened electrostatic repulsion and the Asakura-Oosawa model of depletion. We briefly introduce each component of the OCES potential and proceed to examine the behavior of interacting helical filament pairs.

Net attractive interactions between charged, semi-flexible macroions mediated by neutral ``depletants"  has been studied extensively in the context of osmotically-driven bundling biological filaments, such as DNA, f-actin or microtubules\cite{hosek04,lau2009,needleman04}. Newly developed experimental methods that combined high-resolution microscopy by single-molecule force spectroscopy\cite{streichfuss2011} are directly and quantitatively probing these interactions at a pairwise level.  These methods, along with the independent experimental control over repulsive and attractive forces via respective ionic and osmotic solution conditions, make this an ideal class of interactions for the study of inter-filament cohesion.   

We model electrostatic interactions between (hollow) tubular filaments possessing a uniformly charged surfaces, of areal charge density $\rho$, in an aqueous medium at finite concentration of monovalent salts. We model screened electrostatic repulsion between charged surface elements of the Debye-H\"uckel\cite{safran1994} form.  As in Sec.~\ref{sec:interactions_LJ}, we find the total pair-wise electrostatic interaction between opposing filaments given by 
\begin{equation} 
U_{DH} (\theta, \Delta) = \int dA_+ \int dA_- ~u_{DH}\big(|{\bf R}_+- {\bf R}_- |\big) .
\label{eq: dh}
\end{equation}
where 
\be
u_{DH}(r)=\Gamma \frac{ e^{-\kappa r} }{r},
\label{eq:UE_dh}
\ee
with $\kappa=\lambda^{-1}$ the inverse screening length and $\Gamma=\rho^2 \ell_B k_B T$, with $\ell_B = e^2/ (\epsilon k_B T)$ the Bjerrum length (where we neglect dielectric contrast between filaments and solution).

\begin{figure}
\begin{center}
\includegraphics[scale=0.3]{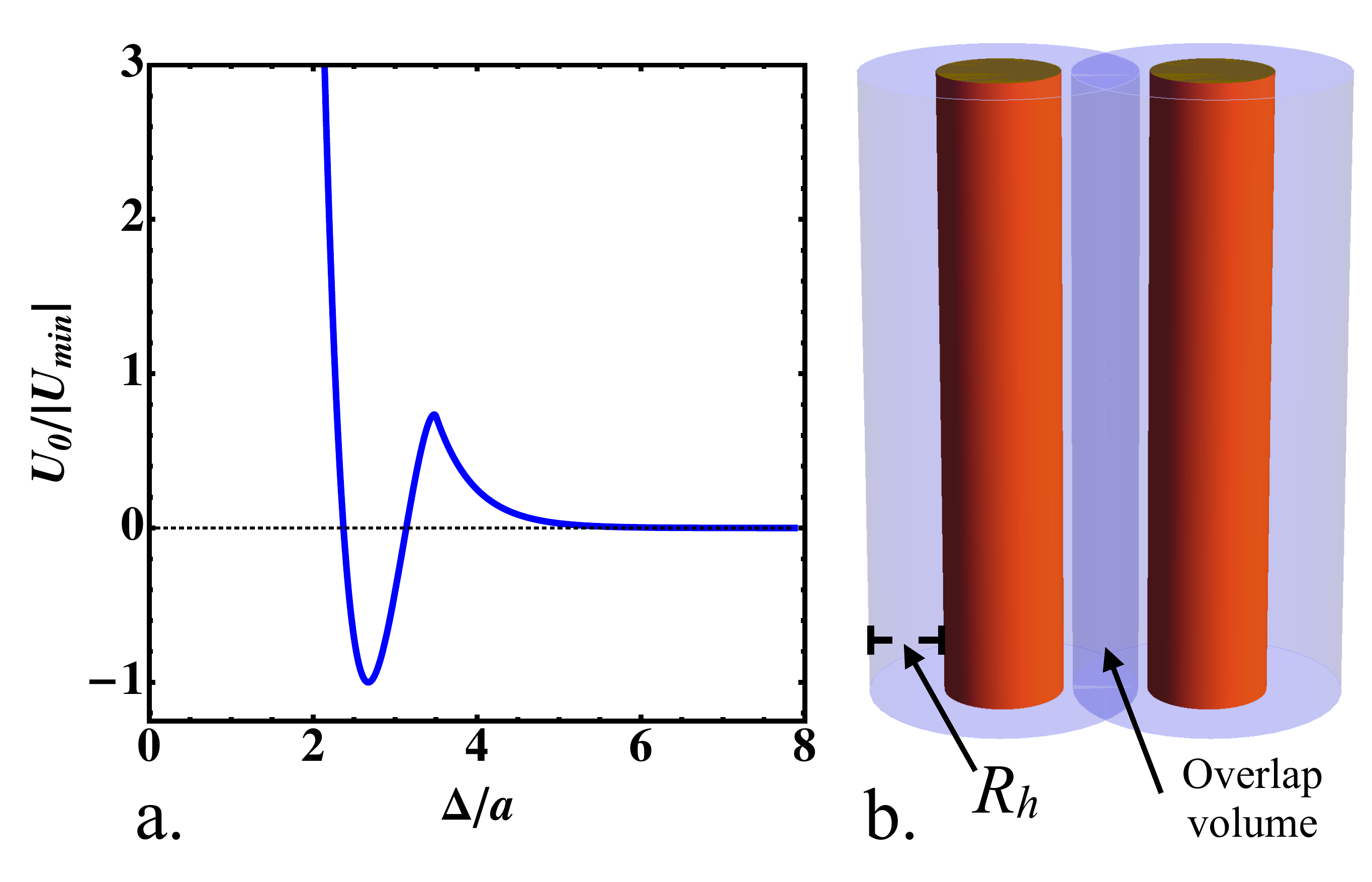}
\caption{(a) Schematic of the interaction energy per unit length for the OCES potential as a function of dimensionless separation $\Delta/a$ for straight filaments ($\theta=0$) with $\lambda/a=0.5$, $R_h/a=0.75$ and $\gamma=1.0$. Notice the long range repulsion part of the potential. (b) Schematic of the halo around the surface of the the filaments for the same $R_h/a$, highlighting the overlap volume in dark purple.}
\label{fig:potential_SED}
\end{center}
\end{figure}

We assume filaments to be immersed in a solution of lower-molecular weight depletants of concentration $c$ whose exclusion from the occupied volume of the much larger filaments induces an osmotic driving force for inter-filament contact, e.g. the {\it depletion effect}.  These entropic attractions are modeled by the Asakura-Oosawa\cite{oosawa1958} theory, where the entropy of hard sphere depletants of radius $R_h$ is determined by the volume $V_{ex}$ of the region excluded to the depletant centers by the filaments.  This volume is enclosed within a surface, pictorially a ``halo'' around the curved filaments\cite{snir2005,snir2007}, that sits a distance $R_h$ from the tubular surface of the filaments. For separated and non-compact configurations filaments, $V^0_{ex} = 2 L \pi (a+R_h)^2$ (neglecting effects from the end).  When the filaments are sufficiently close, the exclusion regions overlap leading to a decrease in $V_{ex}$ by $\Delta V = V^0_{ex}-V_{ex}$.  The entropic free energy gain due to the chance in accessible volume to the depletants $\Delta V$ is simply $-\Pi\Delta V$, where $\Pi$ is the osmotic pressure exerted by the depletant molecules which for sufficiently low concentrations follow the van't Hoff's law $\Pi = c k_B T$.   Given the double helical geometry of skewed filaments, we compute this overlap volume $\Delta V = \ell \Delta A$ in terms of the length of vertical contact $\ell = L \cos \theta$ and the overlap area $\Delta A$ of excluded areas within planar cross sections perpendicular to the pitch axis (see Fig.~\ref{fig:bananas} in Appendix~\ref{app:depletion}).  Given the form of $\Delta A(\theta)$ the depletion interaction per unit filament length is given by 
\be
U_D(\theta)/L= - \Pi \cos\theta \Delta A(\theta).
\label{eq:depletion}
\ee 
Details regarding the calculation of overlap area on skew-angle, $a +R_h$ and $\Delta$ are described in the Appendix~\ref{app:depletion}.  

Two key properties distinguish the nature of depletion forces from either of the other two surface interactions considered in this study (LJ-type and screened electrostatic).  Firstly, depletion interactions are strictly finite range since $\Delta V = 0$ when the filament surfaces are more distant than $2 R_h$, whereas the magnitudes of screened-electrostatic and van der Waals forces decay respectively exponentially or algebraically at large separation.  The second, and more critical, distinction is the non-pairwise additive nature of depletion forces:  the overlap in excluded volume $\Delta V$ cannot be decomposed into a super-position of overlaps per pair of surface elements on opposing filaments.  We show below that the non-pairwise additive nature of depletions gives rise to a qualitatively distinct skew dependence of filament interactions, particularly at small twist.  

An example case of the combined OCES potential $U (\theta) =U_{DH} (\theta) + U_D (\theta)$ is shown in. Fig.~\ref{fig:potential_SED} for parallel filaments ($\theta =0$) for a case where $R_h \approx \lambda$, which exhibits a long-range repulsive/short-range attractive behavior due to the finite range of depletion attraction.  

\subsection{Skew dependence OCES interactions}
\label{subsec:SED}
While the skew dependence of the LJ interactions discussed in Sec.\ref{sec:interactions_LJ} can be classified purely in terms of a single ratio of length scales, the behavior of the OCES interactions, which is governed by interactions of two independent energy and length scales, is not determined by a single dimensionless parameter.  Instead, three dimensionless quantities characterize the skew dependence of the OCES model.  These include both the reduced range of depletion attraction, $R_h/a$, the ratio of attraction to repulsive interaction range, $R_h/\lambda$, as well as the ratio of the magnitudes screened-electrostatic interaction to depletion-induced binding, which can be parameterized by the dimensionless ratio
\begin{equation}
\gamma \equiv \Pi/\Gamma = \frac{c}{ \rho^2 \ell_B}
\end{equation}
where we assume dilute conditions where $\Pi \propto c$.

The reduced potential as a function of $\theta$ is shown in Fig.~\ref{fig:oces} for a series of interaction ranges $R_h/a$ and fixed ratios of attractive to repulsive range $R_h/\lambda$ and of attractive to repulsive strength $\gamma$.   As shown most clearly in Fig.~\ref{fig:oces}(a) for $R_h/\lambda =1$ and $\gamma = 4$, a generic feature of the OCES potential is the appearance of a local maximum separating parallel filaments and large skew angles.  We find that the scaled height of the energy barrier apparently increases with the ratio of interaction range relative to filament radius (below we demonstrate that the small-angle behavior is specifically governed by the electrostatic contributions).  In this subsection, we restrict the discussion to the dependence of the global angle-dependence of $U(\theta)/U_0$ on the interaction range $R_h/a$ and $\gamma$ and reserve a more detailed analysis of the small-$\theta$ behavior for Sec.\ref{subsec:stability_oces}.

Generically, we observe that in the case of very short-ranged interactions where $R_h\ll a$ (for fixed $R_h\approx \lambda$), the reduced binding $U(\theta)/U_0$ approaches the {\it universal} ``sticky tube" limit obtained for $\sigma \ll a$ in Sec.~\ref{subsec:interactions_LJ}.  As in the case of LJ-surface interactions, when interactions are very short-ranged, inter-filament attraction favors maximal surface-surface contact, which is constrained by the {\it universal} geometry of hard-tube packing described in Sec.~\ref{sec:model}.  Hence, in the limit of short-ranged surface interactions, we observe a relative flattening of $U(\theta)/U_0$ for small $\theta$, where the number of line-contacts between tubes remains constant until a critical angle of $\theta_c = \pi/4$ is exceeded, at which point the tubes are able to develop two lines of contact accounting for the limit $U(\theta \to \pi/2) = -2 U_0$.  As $R_h/a \to 0$, we find again that the optimal cohesive angle approaches this critical angle for hard tube packing.   

While limiting behavior of $R_h/a \to 0$ of the OCES conforms the geometric behavior of the ``sticky tube" limit, critical differences between the OCES model and pair-wise LJ model emerge as the interaction range of attraction approaches the tube size, in this case when $R_h \approx a$.  Significantly, we observe that the shift in $\theta_m$ to larger skew angles with increasing interaction range is not a continuous function of $R_h/a$ (see Fig.~\ref{fig:oces_theta_min}).  In particular, for a given value of $\gamma$, we observe for small $R_h/a$ $\theta_m$ increases gradually from $\pi/4$ until a critical value of $R_h/a$ above which $\theta_m$ jumps discontinuously to a maximal twist of $\theta_m = \pi/2$.  Because the helical radius must diverge to avoid inter-tube overlap at maximal twist angle (i.e. $\Delta_* \to \infty$ as $\theta \to \pi/2$), the shift of optimal angle to $\pi/2$  implies that for sufficiently large range of depletion, the optimal cohesion no longer favors a finite filament curvature, which is unlike the thread-like limit $LJ$ potential where we find $\theta_m$ approaches a universal value ($\simeq 58^\circ$) as $\sigma/ a \to \infty$. 

Underlying the loss of optimal angle at intermediate twist angles for $R_h/a \gtrsim 1$ is the geometry of the ``depletion halo" at large twist.  As we show in Appendix~\ref{app:depletion}, for $R_h \gtrsim a$, the ``depletion halos" of opposing filaments are strongly overlapping over the full range of $\theta$, which implies, in this limit, that the angle dependence of  $\Delta A$ is far less sensitive to the precise geometry of inter-filament surface contact than in case of ``sticky tubes", where cohesive contact is maximally sensitive to the non-linear dependence of inter-filament contact on $\theta$.  Specifically, when $R_h/a\gg 1$, as $\theta$ increases, depletion halos begin to make additional contact and overlap with neighboring filaments ``above" and ``below" at skews far below the point where hard tube interactions force in-place spacing $\Delta$ to separate significantly.  Relative to the case of ``sticky tubes" where hard-tube repulsion forces filaments to break cohesive contact with points at equal height for $\theta \gtrsim \pi/4$, large halos maintain substantial in-plane contact (high overlap areas), while simultaneously strengthening attraction with out-of-plane contacts. Hence, for large depletion halos $R_h/a \geq 1$, attractive interactions grow monotonically with twist angle, as we show explicitly in the Appendix, leading to a loss of minimum of $U(\theta)$ for $\theta< \pi/2$.  

In Fig.~\ref{fig:oces}(b) and (c), we demonstrate the effect of increasing $\gamma$ or $R_h/\lambda$, respectively, on the on angle-dependence of inter-filament cohesion.  These examples show that either increasing the {\it magnitude} or {\it range} of depletion relative to screened electrostatic interaction have the same qualitative effect of ``flattening" the potential at small and large $\theta$ 
as indicative of a ``depletion dominated" interactions (see Fig.~\ref{fig:mustache} in Appendix~\ref{app:depletion}).  Notably, as either $\gamma$ or $R_h/\lambda$ increase, the critical interaction range $R_h/a$ at which the optimal adhesion jumps from skewed ($\theta_m< \pi/2$) to fully tilted ($\theta_m = \pi/2$) shifts to larger values, highlighting the role played by electrostatic repulsion in destabilizing filament cohesion at a finite degree of skew.

\begin{figure}
\begin{center}
\includegraphics[scale=0.35]{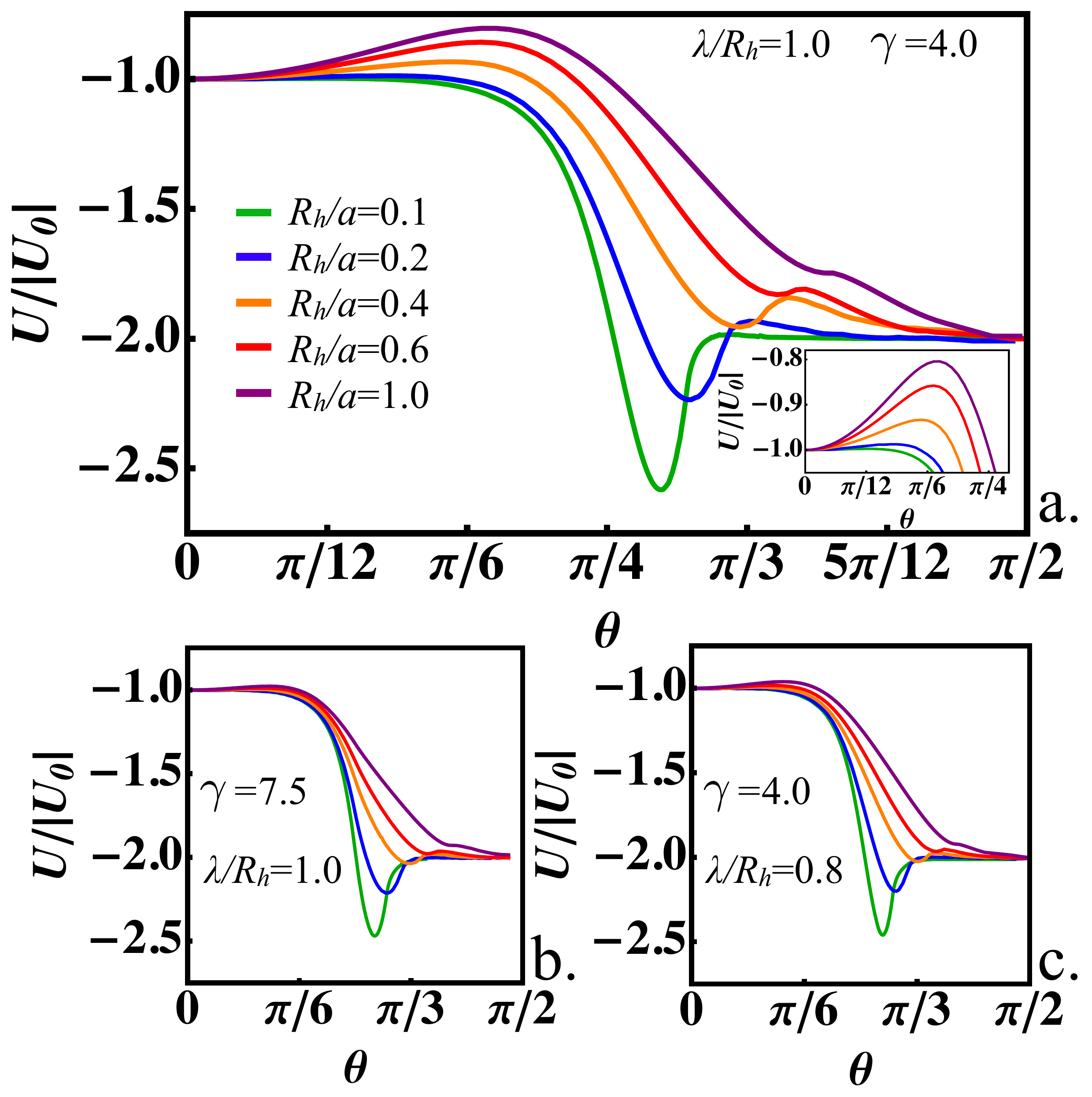}
\caption{(a) Dimensionless energy $U/|U_0|$ as a function of helical angle $\theta$, optimized with respect to $\Delta$, for a range of values of $R_h/a$ for $\lambda/R_h$=1.0 and $\gamma$=4. \emph{Inset:} zoom of the same plot for small helical angles showing an increasing energy barrier with $R_h/a$. (b) Same plot as (a) for a higher $\gamma=7.5$. All curves flatten for small helical angles and the depth of the minima reduces. (c) Same plot as (a) with a smaller ratio of $\lambda/R_h$. The overall trend is the same as that observed in (b).}
\label{fig:oces}
\end{center}
\end{figure}

\subsection{Stability of straight filaments and small-$\theta$ dependence}
\label{subsec:stability_oces}
We now turn to analyze the small-$\theta$ behavior of the OCES potential.  As highlighted in the inset of Fig.~\ref{fig:oces}(a), a universal feature of the OCES behavior is the appearance of a local minimum for $\theta=0$, the stability of cohesive interactions for straight filament pairs $d^2 U(\theta)/d \theta^2|_{\theta=0} = U_2> 0$, and a finite energy barrier separating parallel filaments from lower-energy states at lower skew.  The local minimum at $\theta=0$ is markedly different from the case of LJ-type surface interactions, where parallel filaments are generically unstable to infinitesimal skew.  In this case, the stability of parallel filaments derives from the distinct small-$\theta$ behavior of the components of the OCES interactions.  As discussed above in Sec.~\ref{subsec:SED}, for pairwise interactions the effect of infinitesimal skew can be understood in terms of the increase of the number of points brought into closer interaction range.  For pairwise electrostatic interactions, which are repulsive at all distances, decreasing separation between points relative to parallel filaments implies an {\it increasing} electrostatic repulsion at small-$\theta$, and $d^2 U_{DH}/d\theta^2|_{\theta=0} >0$.  
\begin{figure}
\begin{center}
\includegraphics[scale=0.3]{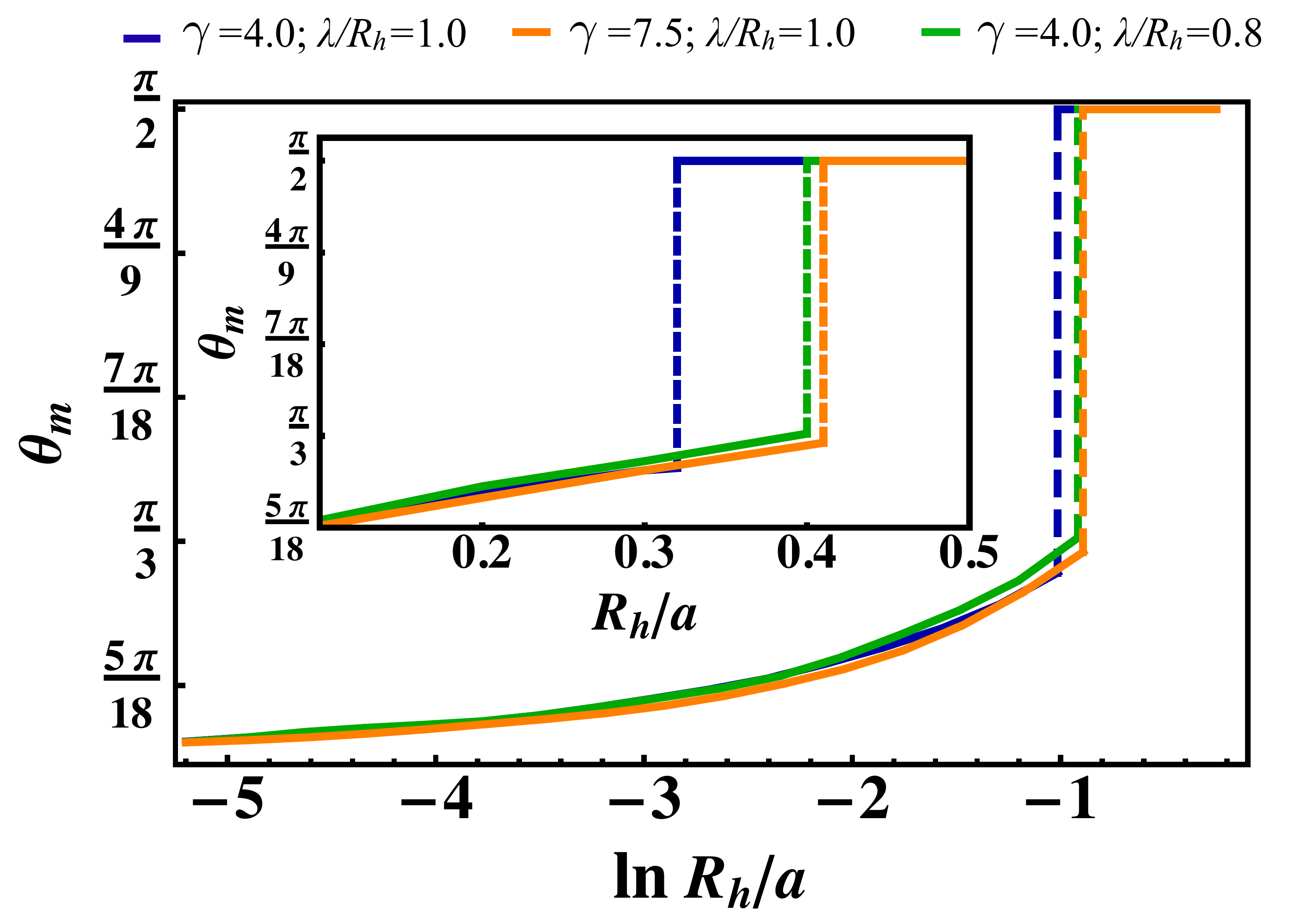}
\caption{$\theta_{m}$ vs. $\ln R_h/a$ for the three cases shown in Fig.~\ref{fig:oces}. \emph{Inset:} The discontinuous jump in the minimum angle shown as a function of $R_h/a$. The discontinuity corresponds to $R_h/a \approx$ 0.32 (blue curve), 0.41 (orange curve) and 0.40 (green curve).}
\label{fig:oces_theta_min}
\end{center}
\end{figure}

Depletion interactions, though they may act at longer range for $R_h  > \lambda$, are not pairwise additive, and therefore, the decreased distance between pairwise elements {\it does not} imply that infinitesimal skew increases depletion induced binding at $O(\theta^2)$.  Instead, it is necessary to consider the change in overlap geometry of the depletion halos surrounding opposing filaments, whereby the effect of twist relates to the increased sectional areas derived from backbone tilt.  For small skew, tubular intersections with the xy plane can be approximated as elliptical cuts through skewed cylinders~\footnote{This follows from the fact that to lowest order in $\Omega$, the backbone curves of filaments approximate straight lines, tilted by $\theta$ with respect to $\hat{z}$.}. Relative to the circular cuts of parallel filaments, at small-$\theta$ elliptical sections are affinely stretched along the tilt direction by a factor of $\sec \theta$, leading to an increase in overlap area $\Delta A(\theta)/\Delta A(\theta=0) \simeq \sec \theta + O(\theta^4)$.  Inserting this asymptotic result into eq.~\eqref{eq:depletion} we find that the geometric increase in overlap area per planar section is {\it perfectly} canceled by the geometric decrease in contact length ($\ell = L \cos \theta$), such that depletion interactions are strictly independent of skew-angle to $O(\theta^4)$ and $d^2 U_{D}/d\theta^2 |_{\theta=0}=0$.   

The insensitivity of depletion for small twist implies that the {\it electrostatic repulsion} always dominates at sufficiently small angles.  Weakly twisting a filament pair generically increases the net free energy, corresponding to a positive curvature of the angle-dependent potential $U_2 = d^2 U_{DH}/d\theta^2|_{\theta=0} >0$.  This implies that parallel filaments ($\theta =0$) are always a weakly metastable state of binding for OCES interactions, and an energy barrier always separates the metastable state of interactions at $\theta=0$ from its minimal value at $\theta=\theta_m$.  Because depletion interactions do not contribute to the potential at $O(\theta^2)$, the stability of parallel filament interactions, as measured through the curvature of the potential $U_2$, is determined solely by the strength and range of electrostatic repulsion.  Hence the inset of Fig.~\ref{fig:oces}(a) shows the curvature and magnitude of the energy barrier to increase with both the range of repulsive interaction (increased $\lambda/a$) and the relative magnitude of electrostatic interactions compared to depletion (decreasing $\gamma$).

\begin{figure}
\begin{center}
\includegraphics[scale=0.26]{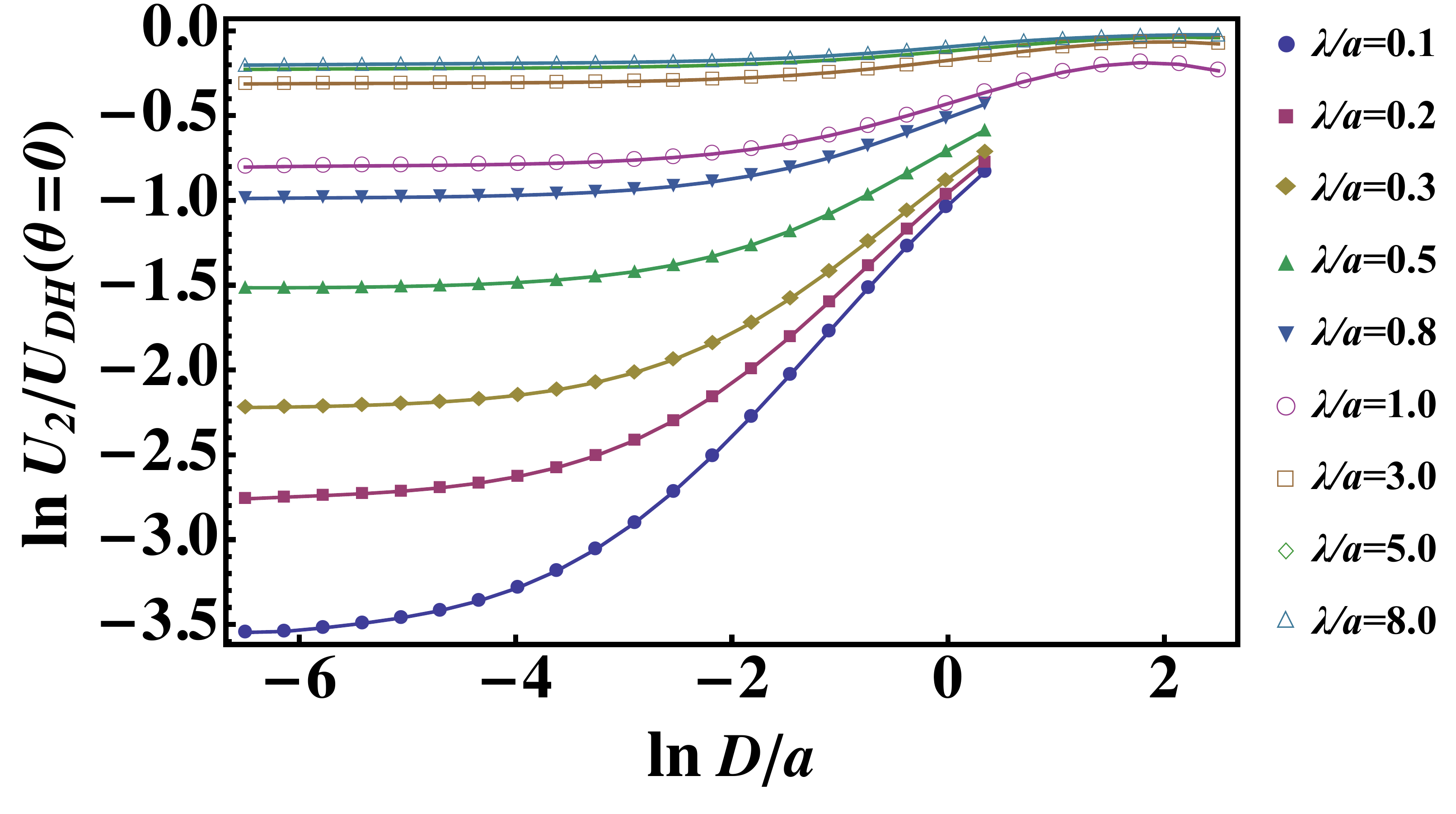}
\caption{Normalized curvature on the potential $\ln U_2/U_{DH}(\theta=0)$ vs. $\ln D/a$, where $D=2(\Delta-a)$ is the surface-surface separation, for various $\lambda/a$. In this plot $\gamma=20$ and $R_h/a=1$.}
\label{fig:U2_plot_v2}
\end{center}
\end{figure} 

In Fig.~\ref{fig:U2_plot_v2} we analyze the ratio $U_2/U_{DH}(\theta =0)$ as a measure of small-twist behavior for a range of ratios $\lambda/a$ and $D/a$, where $D=2(\Delta -a)$ is the surface-surface separation of filaments.  In the well-separated limit, where $D$ is much larger than $a$ and $\lambda$, Fig.~\ref{fig:U2_extra} shows that $U_2/U_{DH}(\theta =0)$ tends towards a characteristic ratio which is largely independent of $\lambda/a$.  Hence, for large $D$, electrostatic sensitivity to twist falls exponentially with separation as $U_2 \sim U_{DH} \sim L \Gamma a^2 K_0(\kappa D)$.  In the limit of close-contact where $D$ is {\it smaller} than both $a$ and $\lambda$ the ratio $U_2/U_{DH}(\theta =0)$ approaches a ratio that is strongly dependent on $\lambda/a$. Fig.~\ref{fig:U2_extra} shows the asymptotic value of $U_2/U_{DH}(\theta =0)$ approached in the close-contact limit $D/a \to 0$.  In this limit, a Derjaguin approximation gives $U_{DH}(\theta =0) = 2 \pi^{3/2} \Gamma a^{1/2} \kappa^{-3/2}$ which implies a scaling behavior of the curvature (for closely-spaced filaments)
\begin{equation}
U_2(D\ll a) /L \sim \Gamma a^2  \times \left\{ \begin{array}{ll} 
(\lambda/a) ^{9/4}  & {\rm for} \ \lambda \ll a 
  \\ \\
 (\lambda/a) ^{3/2} & {\rm for} \lambda \gg a 
 \end{array} \right. .
\end{equation} 
Notably in the limit of short-ranged repulsions $\lambda \ll a$,  the dramatic decrease of potential curvature with decreasing screening length ($U_2 \sim \lambda^{9/4}$) is consistent with ``sticky tube" behavior observed for short interactions ranges in Figs. \ref{fig:oces}, which rapidly flatten in shape as $\lambda/a$ is decreased below unity.


\begin{figure}
\begin{center}
\includegraphics[scale=0.26]{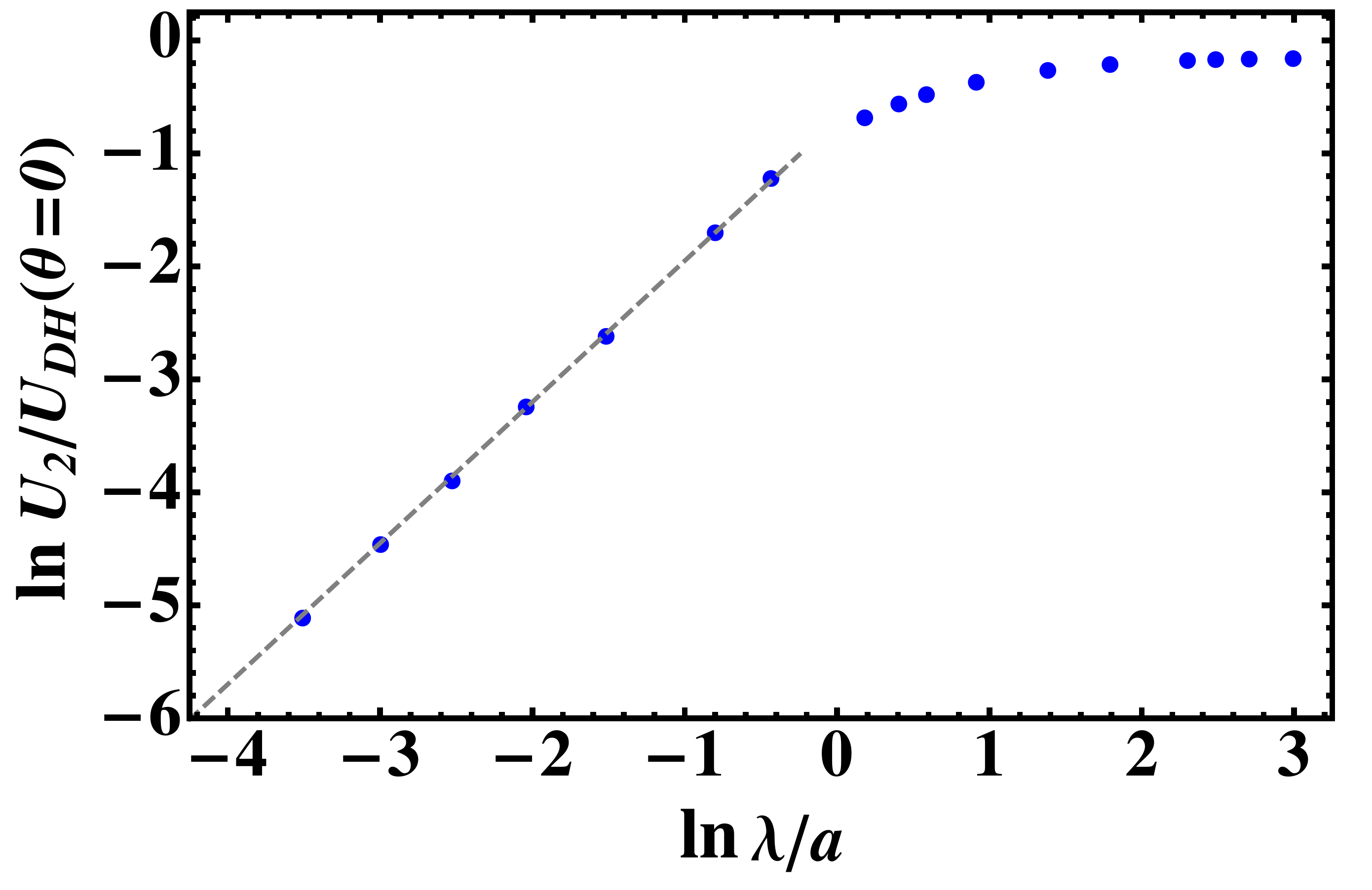}
\caption{Normalized curvature $\ln U_2/U_{DH}(\theta=0)$, as function of $\lambda/a$ in the limit $D/a\to0$ ($\ln D/a\approx -6.14$). In this region $\lambda/a$ scales as a power law $\sim(\lambda/a)^{5/4}$.}
\label{fig:U2_extra}
\end{center}
\end{figure}
\section{DISCUSSION}
\label{sec:disc}
The study of the twist-dependence of inter-filament cohesion yields a number of important general conclusions.  Firstly, that cohesive interactions generically favor non-zero skew.  We can rationalize this in the extreme limit of short-range (``sticky tube") interactions as a geometrical condition for maximizing the number of ``contact lines" between opposing filaments, which corresponds to the crude picture of three lines of contact ``merging" at the critical angle $\theta_c = \pi/4$.   However, the nature of the twist dependence at small angles is demonstrated to be  non-universal, with the sign of the curvature of $U(\theta)$ (and the stability of parallel filaments) determined critically by the nature of interactions dominating in this regime.  For pair-wise surface forces, infinitesimal twist of parallel filaments, upon aggregation, {\it decreases} the distance between points on opposing filaments.   Thus, weak twist generically decreases (increases) the inter-filament potential for attractive (repulsive) interactions relative to the parallel configuration.  When attractive interactions dominate at small twist as in the LJ potential, inter-filament cohesion destabilizes parallel filaments, while the dominance of electrostatic repulsion (combined with the insensitivity of depletion to small twist) leads to meta-stable parallel state.

It is critical to recognize that these conclusions, even at a qualitative level, are critically sensitive to the ability of flexible filaments to conform to a curved, double-helical geometry.  Indeed, for the extreme case of perfectly rigid filaments well-studied in the literature\cite{adrian1972,adrian1974,fixman1978}, precisely the {\it opposite} behavior is achieved at small-twist due to the loss of inter-filament contact at distant ends of the filaments (see Fig.~\ref{fig:double_helix}(a)) and twisting parallel rigid filaments decreases (increases) the potential for  attractive (repulsive) pair-wise interactions between surface elements. For example, if one assumes the twist dependence of electrostatic interactions between {\it rigid cylinders}~\cite{adrian1974, fixman1978} to describe assemblies of flexible, double helically-twisted filaments then one arrives at the conclusion that electrostatic repulsion favors inter-filament helical skew and spontaneous twist of the assembly, a conclusion clearly contradicted by the present analyses. However, such generic conclusions strictly apply only when the center line between double-helically pairs is {\it straight}. The skew-dependence of interactions is likely to be altered in magnitude, and potentially even sign, as the center line between double-helical pair becomes {\it highly curved}, a motif that occurs for certain pairs in large multi-filament assemblies.

We briefly discuss implications of our 2-filament study for models of cohesive assembly of $N>2$ filaments.  We divide this discussion into two prototypical assembly geometries:  i) twisted filament bundles, which are roughly isotropic in cross-section~\cite{grason2007}; and ii) helical filament ribbons, which are highly anisotropic in cross-section, approximately 2D surfaces composed of 1D lateral assemblies of twisted filaments and which have recently been studied as structural models of twisted amyloid fibrils~\cite{lara2011,adamcik2011,mezzenga2013}.  First it is important to note that in both classes of $N>2$ assembly, filaments (or ``strands") belong to a common class of shapes: helices of constant pitch, with helical radii that varies with position in the cross-section (distance from the center).  Furthermore, locally each filament pair has a similar topology to the double-helical geometry of the present study:   filament pairs wind around another with a non-zero, inter-filament skew angle dependent on radial separation.  While, it is therefore natural to expect many of the generic features of the skew-dependence of double-helical pairs to carry over to the pair-wise interactions between constituent filaments in $N>2$ assemblies, intuitively similar twisted geometries are not necessarily described by the same cohesive behavior, as evidenced by the difference between rigid and double-helically twisted filament pairs.  Notably, for bundles (case i) filament separations along azimuthal directions in the bundle are {\it geometrically frustrated} by twist, which significantly complicates the structure and energetics of $N\gg 2$ bundles, as has been studied in detail by Bruss and Grason~\cite{bruss2012,bruss2013}.  Notwithstanding complexities associated with changes in packing topology with twist, it was shown for a model of ``thread-like" LJ-like interactions that such cohesive interactions ultimately favor spontaneous twist of sufficiently long and flexible filament bundles, consistent with the overall effect of such interactions at the pair-wise level of the double helix~\cite{bruss2013}.   

For the somewhat simpler geometry of helical ribbons (case ii), which might be viewed as a single ``row" of filaments in the twisted bundle and where azimuthal frustration is not present, the generic conclusions of the present study of double helices (i.e. that pair-wise surface interactions generically increase in magnitude with twist) can only be rigorously applied for filament pairs sufficiently close to the center of the ribbon, whose geometry is well-approximated by the straight double helix.  Filament pairs at the outer edge of helical ribbons are not only locally twisted, but the contact line between them is also bent (into a helix)~\cite{adamcik2011}.  While a careful analysis of the simultaneous effects of twist and bend cohesion microscopic models of cohesion is beyond the scope of the present study, there is reason to suspect that this bending generically leads to the opposite effect as double-helical twist, due to a ``slip'' of inter-filament contact required at the ends of the ribbon.  Such an effect has been studied in detail for a model of thread-like cohesion in Bruss and Grason~\cite{bruss2013} and has been shown to decrease the magnitude of pair-wise forces with increased twist.  We therefore speculate that the skew-dependence of cohesion will change at least in magnitude, if not sign, as filament pairs become sufficiently far from the center of the ribbon (i.e. for sufficiently wide ribbons).  

Returning to the case of an attractive, pair-wise interaction between filaments, such as the LJ potential, we are led to the surprising and generic conclusion that the cohesive gain (growing as $\sim \theta^2$) dominates the mechanical cost of helical bending (growing as $\sim \theta^4$) for small twist, and hence, cohesion always stabilizes some measure of inter-filament twist.  Given that LJ-type interactions are a common model for the van der Waals interactions between a broad range of neutral filaments, most notably carbon nanotubes, we are led to ask about the magnitude of the equilibrium twist angle predicted by the analysis and compare this to existing experimental observations.   Here, we provide a simple estimate for the equilibrium inter-filament skew of 2-filament carbon nanotube ropes considering (6,6) and (20,20) nanotubes as limiting cases of small- and large-diameters.   For dispersion interactions between graphene-like surfaces, interactions parameters have been computed $\epsilon = 3.48 ~ {\rm eV}/{\rm nm}^4$\cite{girifalco2000} and $\sigma = 0.38 ~ {\rm nm}$\cite{liang2005}, while the bending stiffnesses have been predicted to be $B_{(6,6)} \simeq 712 ~ {\rm eVnm}$ and $B_{(20,20)} \simeq 22,800 ~ {\rm eV nm}$\cite{liang2005,girifalco2000,arroyo2004}.  Using the scaling estimate of eq.~\eqref{eq:U_2_LJ} and \eqref{eq:theta_eq} we estimate $\theta_{eq}$ to be $0.2^\circ$ and $0.08^\circ$, for (6,6) and (20,20) nanotubes, respectively.  Given the nanometer scale of tube diameters, these extremely modest degrees of twist (deriving from the profound rigidity of nanotubes) correspond to helical pitches on the scale of microns (or larger) which would be challenging, at best, to resolve via most the most commonly used microscopy techniques.

\section{Conclusion}
\label{sec:concluding}
In this paper we have demonstrated the importance of flexibility in interacting filaments. Simply put, the large-scale deformation of filaments cannot be ignored when modeling the dependence of filament forces on inter-filament geometry (separation and orientation).  From the point of the inter-filament forces, the double-helical state is ideal given its ability to maintain cohesive contact at the expense of adjusting inter-filament spacing and local geometry.  Considerations of filament mechanics notwithstanding, cohesion favors skewed states, with an optimal geometry determined by the interplay of universal geometric considerations and non-universal features of the inter-filament potential, notably the range, pair-wise vs. non pair-wise additivity, and attractive vs. repulsive nature. Given the markedly distinct predictions of the present model, which assumes filaments to be sufficiently flexible to maintain contact along their contours and the previously studied extreme limit of perfect rigidity, it remains an open challenge to resolve how the twist-dependence of inter-filament cohesion evolves (and ultimately inverts) between these two asymptotic limits for filaments of finite length and stiffness.

\section*{ACKNOWLEDGEMENTS}
\label{sec:thankyou}
The authors would like to thank I. Bruss for thoughtful comments on this manuscript. This work was supported by grant NSF CMMI 10-68852.

\appendix
\section{Depletion interactions between twisted filaments}
\label{app:depletion}

In this appendix, we provide details on the calculation of the depletion-induced attraction in flexible double-helical pairs, eq.~\eqref{eq:depletion}.  Specifically, we compute the overlap area between the filaments in the xy plane, $\Delta A(\theta)$, for a given inter-filament separation $\Delta$ and range of depletion $R_h$~\footnote{In contrast to refs.\cite{snir2005,snir2007} which numerically analyzed vertical cuts of helical tubes, we consider the horizontal cuts, whose shape can be determined analytically} . The excluded volume region is enclosed with in the depletion halo which surrounds the filament and sits at a distance
\begin{equation}
a_{eff} = a + R_h,
\end{equation}
from the center line. To compute the area of overlap between excluded volume regions on opposing tubes, we make use of Green's theorem\cite{arfken2013} for the area bounded by a closed, 2D  curve, ${\bf X(s)}$, in the xy plane\footnote{Green's theorem is a special case in 2D of Stokes theorem} 
\begin{equation}
{\rm Area}\big[{\bf X}(s) \big]=\oint ds~ ({\bf X} \times \partial_s {\bf X} ) \cdot \hat{z}.
\end{equation}
To determine the curves that bound the overlap regions (cross-sectional cuts of the depletion halo with respect to its helical axis) we use the parametrization given by eq.~\eqref{eq: Rpm} 
(under the replacement $a \to a_{eff}$).  Note that integrating the overlap area over the contact height of the pair, $\ell = L \cos \theta$, yields the overlap volume depicted schematically in Fig.~\ref{fig:potential_SED}(b). This involves seeking solutions to the equation ${\bf R_\pm} \cdot \hat z =0$, which via eq.~\eqref{eq:tantheta} may be written as $z^0(\phi_\pm)=-a_{eff} \sin\phi_\pm\sin\theta$. The curves ${\bf R_{\pm}}|_{z^0}\equiv{\bf  R}^0_{\pm}(\phi_\pm)$, valid for any helical angle $\theta$, delimit the boundaries of the cross-sectional area of excluded volume, whose shape is shown in  Fig.~\ref{fig:bananas}.  

The depletion interaction is determined by the area of overlap between the $\pm$ curves.  Let ${\bf C}_\pm$ correspond to the set of $\phi_\pm$ for which ${\bf R}^0_{\pm}$ is overlapping with ${\bf R}^0_{\mp}$.  The sets $\phi_{\pm} \in {\bf C}_\pm$ satisfy the inequalities ${\bf R}^0_{-} (\phi_{-}) \cdot \hat{x} > 0$ and ${\bf R}^0_{+} (\phi_{+}) \cdot \hat{x} < 0$, respectively.  Summing the (signed) area integration over the disjoint overlap contours is equivalent to the sum of contour integrals around the piecewise-continuous curves bounding the overlap areas.  Hence, the overlap area can be expressed as 
\begin{multline}
\label{eq: deltaA}
\Delta A (\theta) =  \oint_{{\bf C}_- } d \phi_- ({\bf R}^0_- \times \partial_{\phi_-} {\bf R}^0_- ) \cdot \hat{z} \\ +\oint_{{\bf C}_+ } d \phi_+ ({\bf R}^0_+ \times \partial_{\phi_+} {\bf R}^0_+ ) \cdot \hat{z}  .
\end{multline}

\begin{figure}[htbp]
\begin{center}
\includegraphics[scale=0.35]{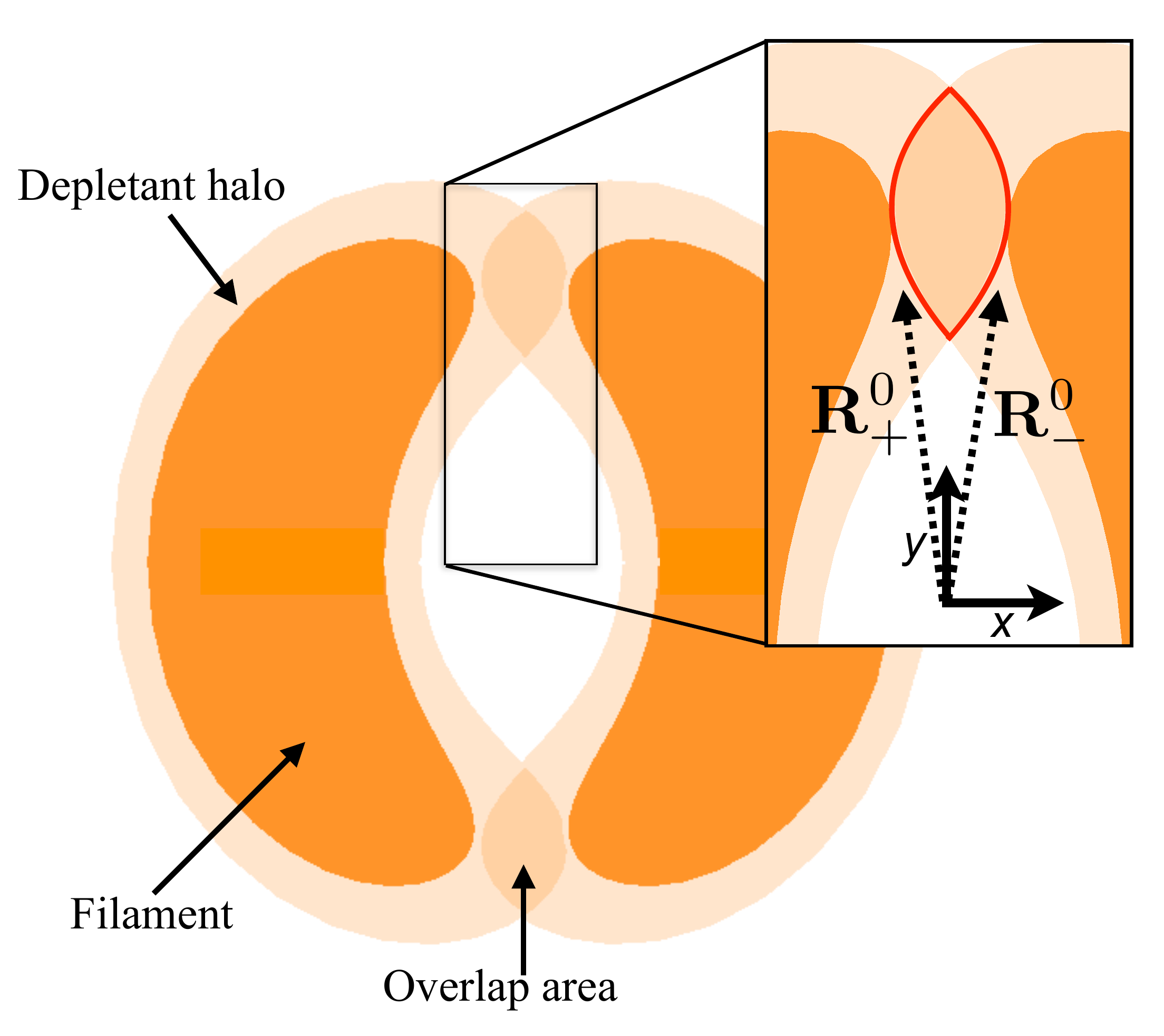}
\caption{A sample cross section of the filaments and depletant halo at non-zero skew ($\theta=7\pi/18$, 
$R_h/a=0.25$ and $\Delta/a\approx2.1$). \emph{Inset:} zoom of the overlap area between the halos showing the origin of the coordinate system used in the parametrization. The red curve corresponds to the region used in the integration in eq.~\eqref{eq: deltaA}}.
\label{fig:bananas}
\end{center}
\end{figure} 


The computed depletion interaction, eq.~\eqref{eq: deltaA}, as a function of helical angle is shown in Fig.~\ref{fig:mustache}, for the limiting case when the filaments are closed packed (the distance of closest approach between center lines is $a$), corresponding to maximal overlap volume. Note the universally (flat) behavior at small helical angles, signature of a low energy gain that increases as a quartic power. Again, the small-$\theta$ dependence can be easily understood by the shape of cross-sections at small angles, which are affinely strecthed by a factor of $1/\cos \theta$ due to small in-plane tilt of the backbone curve. In this limit, eq.~\eqref{eq: deltaA} scales as $\Delta A(\theta=0)/\cos\theta$, perfectly balancing the $\cos\theta$ term of the osmotic attraction, eq.~\eqref{eq:depletion}, thus making the leading power to scale as $\theta^4$. For higher values of $\theta$ the depletion reaches a global minima, which shifts rightward with increasing halo size. In the limit where the depletion halo is negligible, the resulting attraction is dominated by the geometry of the close-packed double-helices described in Sec.~\ref{sec:model}.  Finally, as the helical angle increases, the cross sections have deformed into elongated jelly beans, as shown in Fig.~\ref{fig:bananas}, where there are only two points of contact providing adhesive gain, which accounts for the saturation value of $-2U_D(\theta=0)$ in the energy. 

\begin{figure}
\begin{center}
\includegraphics[scale=0.285]{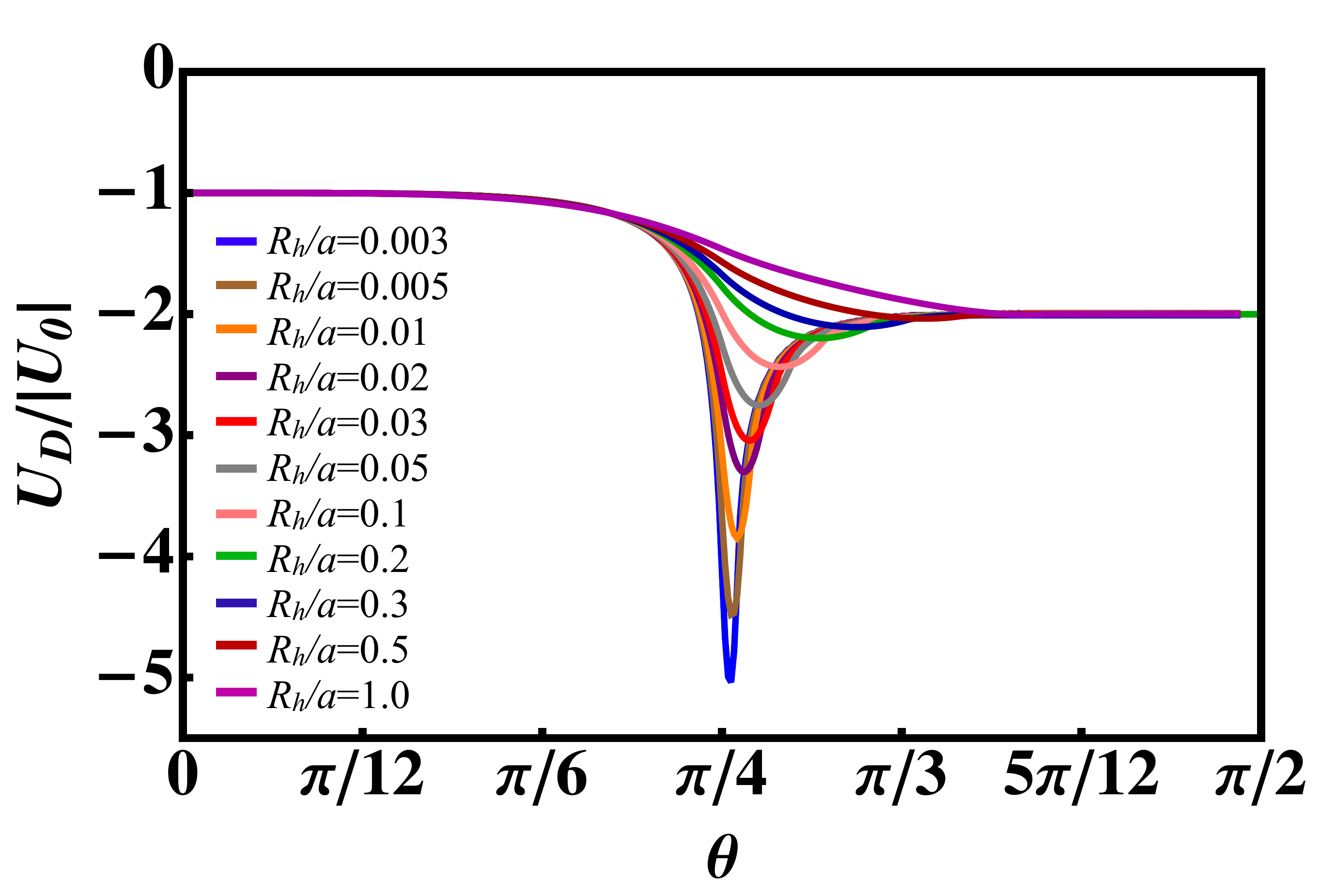}
\caption{Normalized depletion interaction as a function of helical angle. The filaments are always in contact, $\Delta=a$, and the ratio of the depletion halo to radius of filaments, $R_h/a$, changes. The scaling of the energy at small helical angles is $\sim \theta^4$.}
\label{fig:mustache}
\end{center}
\end{figure}

%
\bibliography{bib_v2.bib}

\end{document}